\documentclass[useAMS,usedcolumn,usenatbib,usegraphicx]{mn2e}

\usepackage{epsfig,graphicx,natbib}
\usepackage{./reference/mycite}
\usepackage{amssymb}
\usepackage{amsfonts}
\usepackage{amsmath}
\usepackage{color}
 
\begin{document}
\title[Photo-z for DES and VHS]{Photometric Redshifts for the Dark Energy Survey and VISTA and Implications for Large Scale Structure}
\author[Banerji et al.]
{Manda Banerji$^{1}$ \thanks{E-mail: mbanerji@star.ucl.ac.uk},
Filipe B. Abdalla$^{1}$,
Ofer Lahav$^{1}$, 
Huan Lin$^{2}$ \\
$^{1}$Department of Physics and Astronomy, University College London,
Gower Street, London, WC1E 6BT, UK.\\
$^{2}$Center for Particle Astrophysics, Fermi National Accelerator Laboratory, Batavia, IL 60510. \\ 
}
\maketitle

\begin{abstract}
We conduct a detailed analysis of the photometric redshift requirements for the proposed Dark Energy Survey (DES) using two sets of mock galaxy simulations and an artificial neural network code - ANNz. In particular, we examine how optical photometry in the DES $grizY$ bands can be complemented with near infra-red photometry from the planned VISTA Hemisphere Survey (VHS) in the $JHK_s$ bands. We find that the rms scatter on the photometric redshift estimate over $1<z<2$ is $\sigma_z$=0.2 from DES alone and $\sigma_z$=0.15 from DES+VISTA, i.e. an improvement of more than 30\%. We draw attention to the effects of galaxy formation scenarios such as reddening on the photo-z estimate and using our neural network code, calculate the extinction, $A_v$ for these reddened galaxies. We also look at the impact of using different training sets when calculating photometric redshifts. In particular, we find that using the ongoing DEEP2 and VVDS-Deep spectroscopic surveys to calibrate photometric redshifts for DES, will prove effective. However we need to be aware of uncertainties in the photometric redshift bias that arise when using different training sets as these will translate into errors in the dark energy equation of state parameter, $w$. Furthermore, we show that the neural network error estimate on the photometric redshift may be used to remove outliers from our samples before any kind of cosmological analysis, in particular for large-scale structure experiments. By removing all galaxies with a neural network photo-z error estimate of greater than 0.1 from our DES+VHS sample, we can constrain the galaxy power spectrum out to a redshift of 2 and reduce the fractional error on this power spectrum by $\sim$15-20\% compared to using the entire catalogue. 

Output tables of spectroscopic redshift versus photometric redshift used to produce the results in this paper can be found at $www.star.ucl.ac.uk/\sim mbanerji/DESdata$.      
\end{abstract}


\section{Introduction}

It is now widely accepted that dark energy is responsible for driving the observed acceleration of the Universe. In recent years, measuring and constraining the nature of this dark energy has become a central focus of current studies in cosmology. Several different methods have been developed and shown to probe the nature of dark energy through its effects on the geometry and structure of the Universe e.g. \citet{Riess:98,Hu:WL,Blake:03}. Large-scale sky surveys such as the Sloan Digital Sky Survey have no doubt aided this kind of study \citep{Tegmark:SDSS,Spergel:WMAP3} and more such galaxy surveys are now being planned to exploit the different techniques that will help us better understand the nature of dark energy. 

The proposed Dark Energy Survey is one such experiment. It will use four independent probes namely galaxy clusters, galaxy power spectrum measurements, weak lensing studies and a supernova survey to constrain the nature of dark energy. Each of these methods relies on accurate distance measurements extending over cosmological scales. Given the wealth of data that will be available to us from such surveys, measuring distances and redshifts for all the objects using spectroscopic methods clearly becomes unfeasible. Hence the need for photometric redshifts.

Photometric redshift estimation methods have been around since the 1960s but have undergone a recent revival with proposals for a new generation of large-scale photometric surveys such as the Dark Energy Survey. New algorithms have been developed for photo-z estimation and made available to the community e.g. \citet{Collister:ANNZ, Bolzonella:HyperZ, Benitez:BPZ, Feldmann:ZEBRA, Babbedge:IMPZ}. Furthermore, there is currently a lot of emphasis on optimising the depth and number of bands that will be used to image galaxies in future galaxy surveys so as to obtain accurate photometric redshifts. It is widely known that imaging in more bands can help reduce errors on photometric redshifts but the costs of adding more filters to planned surveys are significant. Given the plethora of data that is now becoming available to us covering the whole range of the electromagnetic bands and a large portion of the observable sky, it is vital that we explore the overlap between different surveys and fully exploit the data sets available to us, in order to achieve the best compromise between cost and science. 

In this paper we analyse the prospect of combining optical data from the Dark Energy Survey (DES) with near infra-red data from the Vista Hemisphere Survey (VHS) in order to obtain accurate photometric redshifts that will help us better constrain the nature of dark energy. We begin with a brief description of these two proposed surveys and a description of the method used to generate mock galaxy samples for both of these surveys. We then proceed to a full photometric redshift analysis of simulated data from these two surveys using artificial neural networks. We assess the impact of reddening on our photometric redshift estimate as well as the effects of removing outliers and using different training sets. In each case, we present results obtained for the optical data from DES only and for optical and near-infra red data from DES and VHS. Finally, we look at the implications of our results for cosmological constraints on dark energy. In particular, we concentrate on the impact of photometric redshift errors on constraints on dark energy using galaxy power spectrum measurements. All magnitudes quoted in this paper are in the AB system.   

\section{The Dark Energy Survey (DES)}

The Dark Energy Survey is a proposed ground-based photometric survey that will image 5000deg$^2$ of the South Galactic Cap in the optical $griz$ bands as well as the $Y$-band. The survey will be carried out using the Blanco 4-m telescope at the Cerro Tololo Inter-American Observatory (CTIO) in Chile. The main objectives of the survey are to extract information on the nature and density of dark energy and dark matter using galaxy clusters, galaxy power spectrum measurements, weak lensing studies and a supernova survey. This will be achieved by measuring redshifts of some 300 million galaxies in the redshift range $0<z<2$, tens of thousands of clusters in the redshift range $0<z<1.1$ and about 2000 Type 1a supernovae in the redshift range $0.3<z<0.75$ \citep{DES:Whitepaper}. Observations will be carried out over 525 nights spread over five years between 2010 and 2014 and when completed, DES will provide a legacy archive of data extending around two magnitudes deeper than the Sloan Digital Sky Survey which is currently the largest existing CCD survey of the Universe by volume. We have estimated the DES volume to be $23.74h^{-3}Gpc^3$ in the range $0<z<2$, about ten times that of the SDSS LRG sample \citep{Blake:CosmoLRG}. This is assuming a 10$\sigma$ AB magnitude limit of $r \lesssim 24$. 

The DES survey area overlaps with that of several other important current and future surveys for example the southern equatorial strip of the Sloan Digital Sky Survey and the South Pole Telescope SZE cluster survey. The entire DES region will also be imaged in the near infra-red bands on two public surveys being conducted on the Visible and Infra-Red Survey Telescope for Astronomy (VISTA) at ESO's Cerro Paranal Observatory in Chile. 

\section{The Vista Hemisphere Survey (VHS)}

Most of the time on the VISTA telescope has been dedicated to large-scale public surveys. Two of these surveys that are relevant to cosmology are the Vista Hemisphere Survey (VHS) and the Vista Kilo-Degree Infra-red Galaxy Survey (VIKING) - \citep{Arnaboldi:VISTA}.
 
The VISTA Hemisphere Survey is a proposed panoramic infra-red survey that will image the entire southern sky ($\sim$20000deg$^2$) in the near infra-red $YJHK_s$ bands when combined with other public surveys. About 40\% of the total VHS time has been dedicated to VHS-DES, a 4500deg$^2$ survey being carried out in the DES region over 125 nights in order to complement the DES optical data with near infra-red data. The initial proposal is for the survey to image in the $JHK_s$ bands with 120s exposure times in each band reaching $10\sigma$ magnitude limits of $J=20.4$,$H=20.0$ and $K_s=19.4$. A second pass may then be obtained with 240s exposures in each of the three NIR filters in order to reach the full-depth required by DES. The VHS-DES survey assumes that $Y$-band photometry will come from the Dark Energy Survey. 

The remaining 500deg$^2$ of the DES area not covered by VHS-DES, will be imaged by VIKING which is a near infra-red survey designed to provide an important complement to the optical KIDS project being carried out on the VST. The details for all these surveys are summarised in Table \ref{tab::surveys}.

\begin{table}
\begin{center}
    \begin{tabular}{|l|c|c|c|}
      \hline
      Survey & Bands & $10\sigma$ magnitude lims & Area \\
      \hline
      DES & $g$ & 24.6 & 5000deg$^2$ \\
          & $r$ & 24.1 &    \\
          & $i$ & 24.3 &   \\
          & $z$ & 23.9  &    \\
          & $Y$ &  21.6 &   \\
      \hline
     VHS-DES  & $J$ & 20.4 & 4500deg$^2$  \\
          & $H$ & 20.0 &  \\
          & $K_s$ & 19.4 &  \\
     \hline
     VIKING & $Z$ & 22.4  & 1500deg$^2$ \\
            & $Y$ & 21.6  & \\
            & $J$ & 21.4  & \\
            & $H$ & 20.8  & \\
            & $K_s$ & 20.5   & \\
     \hline
     KIDS & $u'$ & 24.1 & 1500deg$^2$ \\
          & $g'$ & 24.6 &  \\
           & $r'$ & 24.4 &  \\
          & $i'$ & 23.4 &  \\
\hline
 \end{tabular}    \vspace{2mm}
  \end{center}
\caption{Areas and $10\sigma$ magnitude limits for the surveys discussed in this work. The magnitudes are in the AB system.               \label{tab::surveys}} 
\end{table} 

\section{Simulating Mock Data}

\label{sec:catalogues}

In this work, we have used two sets of mock galaxy samples as simulations of data from DES and VHS. In this section we briefly describe the way in which these data samples are generated. Both catalogues are generated using Monte Carlo methods after assuming relevant redshift, magnitude and type distributions.

\subsection{DES5yr Sample}

The first mock catalogue is that of \citet{DES:photoz} and \citet{Lin:DESsims} - DES5yr hereafter. It adopts the galaxy magnitude-redshift distribution derived from the luminosity functions of \citet{Lin:LF} and \citet{Poli:LF} and a type distribution derived using data from the GOODS/HDF-N field \citep{Capak:HDF-N,Wirth:GOODS-N,Cowie:GOODS-N} and the CWW template SEDs \citep{CWW:templates}. A flux-limited sample is constructed with $0<z<2$ and $20<i<24$. The photometric errors on each object are computed according to the DES $10\sigma$ $griz$ magnitude limits. Note, no attempt is made here to fit a reddening value to each galaxy. 

\subsection{JPL Mock Catalogue}

\begin{figure*}
\begin{center}
\begin{minipage}[c]{1.00\textwidth}
\centering 
\includegraphics[width=8.5cm,angle=0]{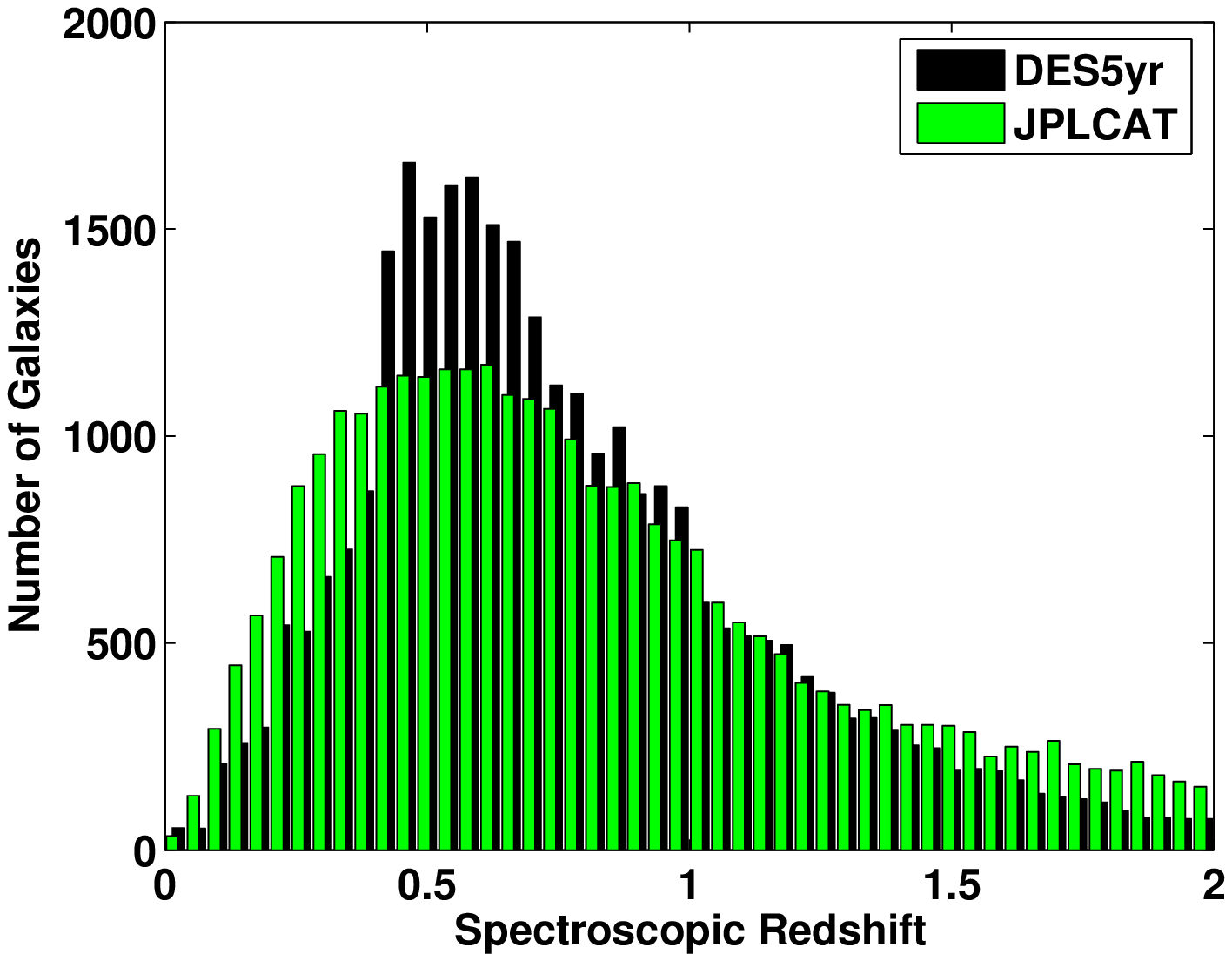}
\includegraphics[width=8.5cm,angle=0]{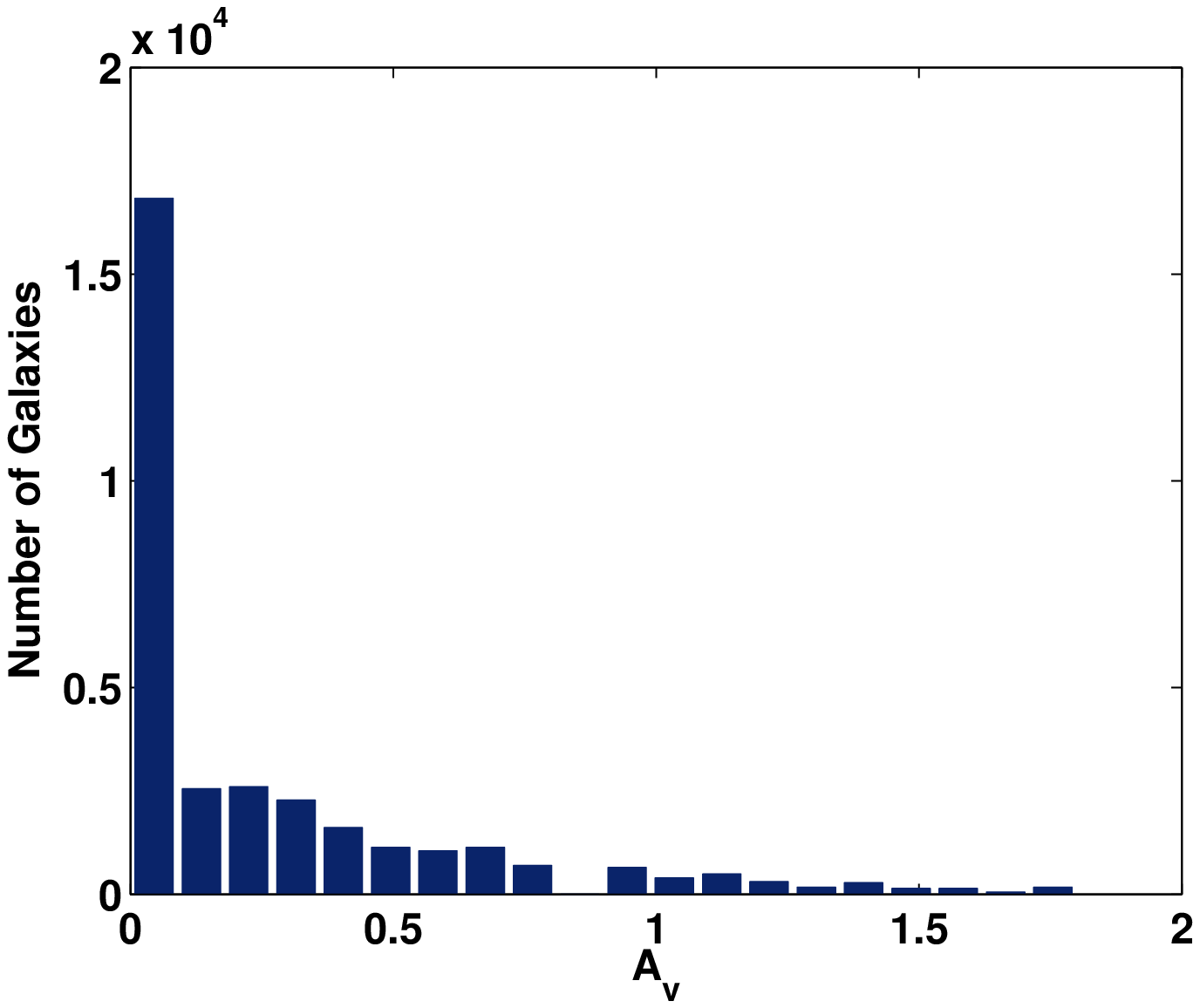}
\end{minipage}
\caption{The distributions characterising our simulated catalogues for the DES and DES+VHS samples. The left-hand panel shows the redshift distribution for $\sim30000$ galaxies in the DES5yr and JPLCAT simulations. All galaxies have $20<i<24$ and $0<z<2$. The right-hand panel shows the distribution of galaxies as a function of the extinction parameter, $A_v$ for the JPLCAT sample.}
\label{fig:cats}
\end{center}
\end{figure*}

The second catalogue is described in \citet{Abdalla:DUNEphotoz} - JPLCAT hereafter. Templates constructed from broadband photometry using a method similar to \citet{Budavari:mockcats} were fit to real objects from the GOODS-N spectroscopic sample \citep{Cowie:GOODS-N, Wirth:GOODS-N} in order to generate this catalogue. The templates were de-reddened and at the time of fitting, the best fit SED and reddening value were found simultaneously. A Calzetti reddening law \citep{Calzetti:red97} was used. Further details of the method used to create photometric data for the catalogue with the correct redshift distribution and luminosity evolution, can be found in $\S2$ of \citet{Abdalla:DUNEphotoz}. For the purposes of this paper, however, the important difference between this and the DES5yr sample is the fact that the galaxies are reddened and corrected for dust extinction. The JPLCAT sample also uses two more templates from \citet{Kinney:temp} to fit the galaxies in addition to the CWW templates used for the DES5yr sample. For the work described in the rest of this paper, the JPLCAT sample when used has been cut so as to have the same magnitude and redshift limits as the DES5yr sample. The redshift distribution for both samples cut to include the same number of galaxies, as well as the distribution of galaxies in the JPLCAT sample for different values of the extinction parameter, $A_v$, are shown in Figure \ref{fig:cats}. \\ \\

Note that the JPLCAT sample is more complex and hence the results from it are likely to be more pessimistic than those for the DES5yr sample. However, both catalogues are generated by fitting models to real data and it is not obvious which of these models captures the true colour variance best. Hence both can be taken as realistic possibilities for modelling the DES and VHS data samples.    

\section{Estimating Photometric Redshifts using Artificial Neural Networks: ANNz}
\label{sec:ANNz}

Methods of estimating photometric redshifts fall into two broad categories, namely template fitting methods and empirical methods. Template fitting methods use libraries of galaxy spectral energy distributions such as the observed Coleman, Wu \& Weedman templates \citep{CWW:templates} or synthetic templates e.g. \citet{BC:popsynth,Fioc:PEGASE}. The spectra are convolved with a filter transmission function in order to calculate the flux through each filter in the filter set being used to observe the object. The fluxes can then be matched to the observed fluxes of different objects using a $\chi^2$ minimisation to output the best-fit redshift and type of the galaxy. Popular photo-z codes that use this method include HyperZ \citep{Bolzonella:HyperZ}, BPZ \citep{Benitez:BPZ} and many others.

Empirical methods on the other hand rely on the availability of a suitably representative training set that can be used to determine the functional relation

\begin{equation}
z = z(\overline m,\overline w)
\label{eq:ztrain}
\end{equation}

\noindent where the redshift is some function of the magnitudes, $\overline m$, and some weights, $ \overline w$.

Once the redshift is known as a function of the magnitude, this relation can be applied to a data set where only the magnitude is known in order to determine the redshift. Examples of this method include polynomial fitting \citep{Connolly:polynomial}, nearest neighbours \citep{Csabai:photoz} and artificial neural networks \citep{Collister:ANNZ} among others. 

Artificial Neural Networks have been shown to produce competitive results compared to other training set methods \citep{Firth:neuralnets} and we use the code ANNz \citep{Collister:ANNZ} to calculate photometric redshifts in all the work that is described in this paper. The neural network is made up of several layers, each consisting of a number of nodes. The first layer receives the galaxy magnitudes in different filters as inputs and the last layer outputs the estimated photometric redshift. All nodes in the hidden layers in between are interconnected and connections between nodes $i$ and $j$ have an associated weight, $w_{ij}$.  

ANNz, like all other empirical methods, requires a training set that is used to minimise the cost function, $E$ (Eq. \ref{eq:cf}) with respect to the free parameters $w_{ij}$. 

\begin{equation}
E=\sum_{k}(z_{phot}(w_{ij},m_{k})-z_{train,k})^2
\label{eq:cf}
\end{equation}

\begin{figure}
\begin{center}
\includegraphics[width=8.5cm,angle=0]{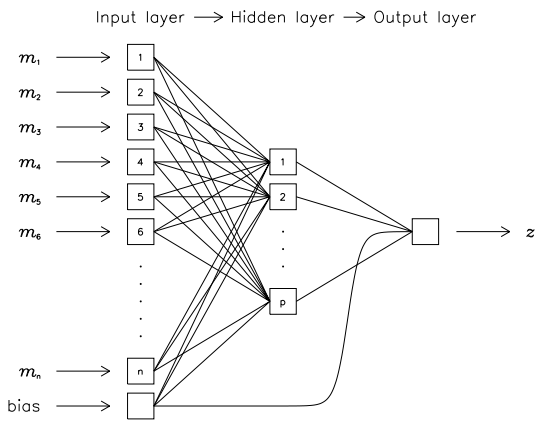}
\caption{Schematic diagram of neural network as implemented by ANNz from \citet{Collister:ANNZ}. The input layer consists of nodes that take  magnitudes in the different filters used for photometry. A single hidden layer consisting of p nodes is shown here although more hidden layers could be used. The output layer has a single node that gives the photometric redshift. Once again further nodes for more outputs such as spectral type could be added to this layer. Each connecting line between nodes carries a weight, $w_{ij}$. The bias node allows for an additive constant when optimising weights.}
\label{fig:ANNz}
\end{center}
\end{figure}

The neural network setup is illustrated in Figure \ref{fig:ANNz}. If the data is noisy, a validation set may be used in addition to the training set to prevent over-fitting. During the initial setup, one has to specify the architecture of the neural network - the number of hidden layers and nodes in each hidden layer. We choose this to be N:2N:2N:1 throughout this work unless otherwise mentioned, where N is the number of filters used for photometry. Note that we have tried changing both the number of hidden layers as well as the number of nodes in each hidden layer and find that it makes very little difference to the photometric redshift estimate.

The neural network code produces an estimate of the error associated with each photometric redshift estimate in addition to the photo-z estimate. This error depends on the noise on the neural network inputs and not on the difference between the spectroscopic and photometric redshifts. The variance that this noise on the input would introduce into the output of the network is given by a simple chain rule expression as follows:

\begin{equation}
{\delta_z}^2=\sum_{i} \left(\frac{\partial z}{\partial m_i}\right)^2
\delta^2_{m_i} 
\label{eq:errANNz}
\end{equation} 

\noindent where the sum $i$ is a sum over all the network inputs and $\delta_{m_i}$ is the photometric error on the magnitude in band $i$. The derivative $\frac{\partial z}{\partial m_i}$ is obtained using the formalism described in \citet{Bishop}. This algorithm is fully implemented within ANNz \citep{Collister:ANNZ}. 

\section{Photometric Redshift Analysis}

\subsection{Choice of Filters}
\label{sec:NIR}

In this section we look at the impact of different filter combinations and survey depths on the photometric redshift estimate. We do this by running the neural network code described in $\S$\ref{sec:ANNz} on the DES5yr sample described in $\S$\ref{sec:catalogues}. ANNz was run on the mock data for five different filter configurations. These are summarised in Table \ref{tab:filters}.

\begin{table}
\begin{center}
    \begin{tabular}{|l|c|c|c|}
      \hline
      No. & Filters & Exposure Time & $10\sigma$ magnitude limits \\
      \hline
      1 & DES $g$ & 400s & 24.6 \\
        & DES $r$ & 400s & 24.1 \\
        & DES $i$ & 1200s & 24.3 \\
        & DES $z$ & 1600s & 23.9 \\
      \hline
      2 & DES $griz$ & As in 1 & As in 1 \\
        & DES $Y$ & 400s & 21.6 \\
      \hline
      3 & DES $griz$ & As in 1 & As in 1 \\
        & VHS $J$ & 120s & 20.4 \\
        & VHS $H$ & 120s & 20.0 \\
        & VHS $K_s$ & 120s & 19.0 \\
      \hline
      4 & DES $griz$ & As in 1 & As in 1 \\
        & DES $Y$ & As in 2 & As in 2 \\
        & VHS $JHK_s$ & As in 3 & As in 3 \\
      \hline
      5 & DES $griz$ & As in 1 & As in 1 \\
        & DES $Y$ & As in 2 & As in 2 \\
        & VHS $J$ & 240s & 21.2 \\
        & VHS $H$ & 240s & 20.8 \\
        & VHS $K_s$ & 240s & 20.2 \\ 
\hline
 \end{tabular}    \vspace{2mm}
  \end{center}
\caption{Summary of filter configurations of DES and VHS considered in \S \ref{sec:NIR}.              \label{tab:filters}} 
\end{table}

 \begin{figure*}
\begin{center}
\begin{minipage}[c]{1.00\textwidth} 
\centering 
\begin{tabular}{cc}
\huge{1} & \huge{2} \\
\includegraphics[width=8.5cm,angle=0]{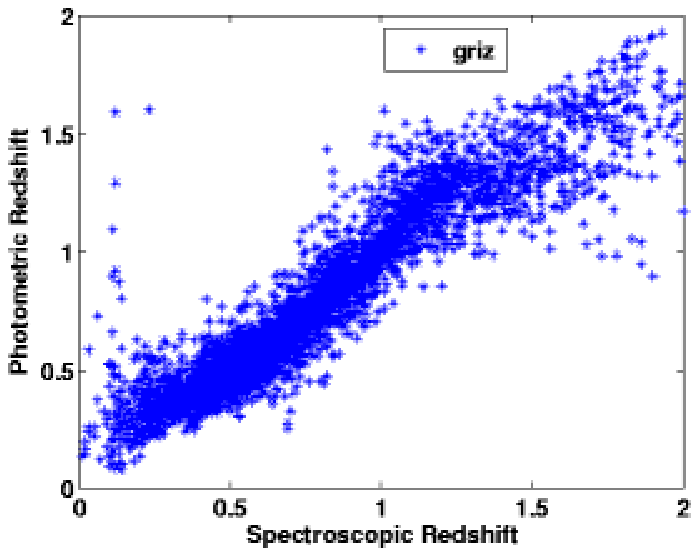} & \includegraphics[width=8.5cm,angle=0]{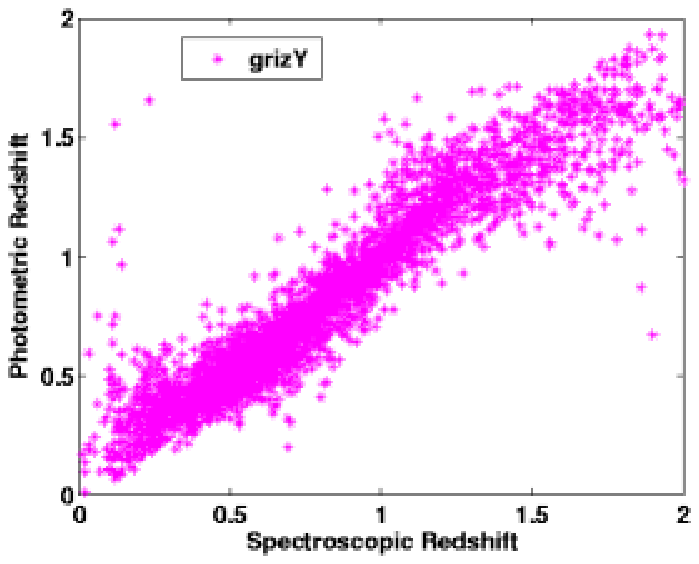} \\
\end{tabular}
\begin{tabular}{c}
\huge{3} \\
\includegraphics[width=8.5cm,angle=0]{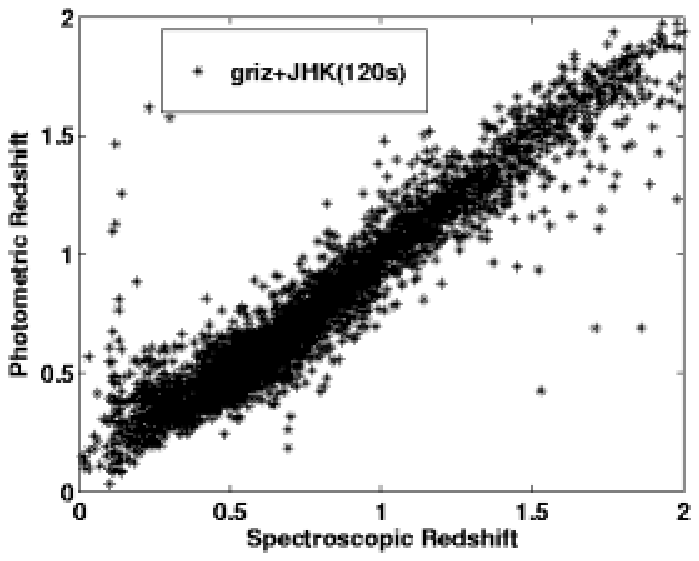}
\end{tabular}
\end{minipage}
\begin{minipage}[c]{1.00\textwidth} 
\centering 
\begin{tabular}{cc}
\huge{4} & \huge{5} \\
\includegraphics[width=8.5cm,angle=0]{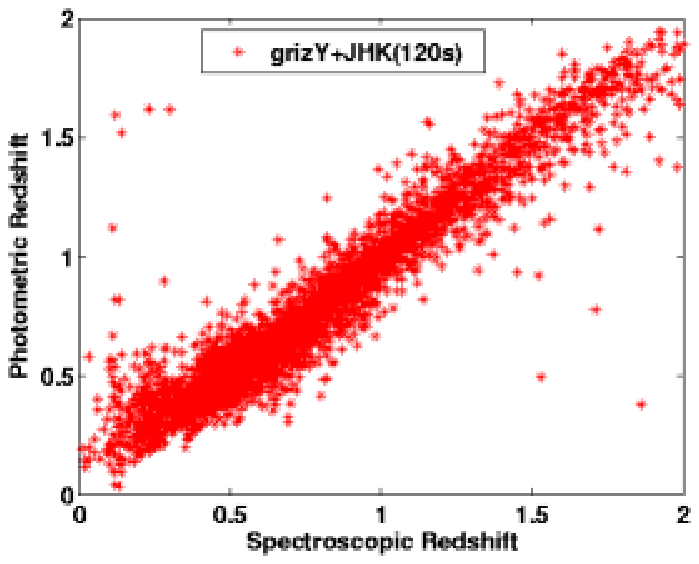} & \includegraphics[width=8.5cm,angle=0]{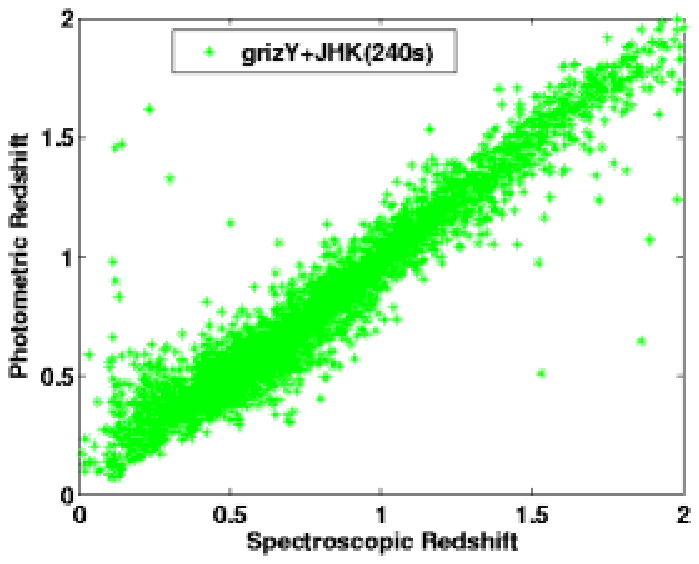} \\
\end{tabular}
\end{minipage}
\caption{Scatter plots of photometric redshifts as a function of the true redshifts for each of the different survey configurations detailed in Table \ref{tab:filters}. These plots are generated for a sample of 5000 galaxies.
\label{fig:DESscatter}}
\end{center}
\end{figure*}

We computed photometric redshifts for each of these cases and from the available true redshifts, computed the scatter on the photo-z estimate. The scatter is the rms photometric redshift error around the mean and is defined in the following way:

\begin{equation}
\sigma_{z}={\left<(z_{spec}-z_{phot})^2\right>}^{\frac{1}{2}}
\label{eq:scatter}
\end{equation}

\noindent where the scatter is evaluated in a redshift bin between $z_1$ and $z_2$. 

\begin{figure}
\begin{center}
\includegraphics[width=8.5cm,angle=0]{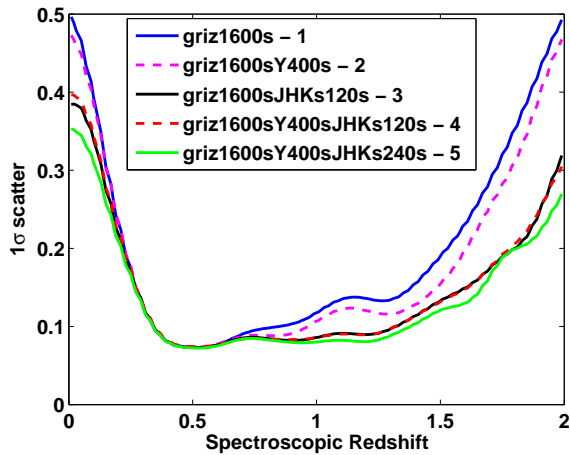}
\caption{The $1\sigma$ scatter on the photometric redshift as a function of the spectroscopic redshift for each of the survey configurations detailed in Table \ref{tab:filters}. Curves are labelled 1 to 5 corresponding to the numbers in Table \ref{tab:filters}.}
\label{fig:DESsig}
\end{center}
\end{figure}

Figure \ref{fig:DESscatter} and Figure \ref{fig:DESsig} show the results of this study. We can see that inclusion of the NIR filters leads to an improvement in $\sigma_{z}$ by $\sim$30\% for $z>1$. $\sigma_z$ is 0.2 for $z>1$ for the DES only sample and 0.15 for the DES+VHS sample in the same redshift range. Increasing the exposure time in the NIR also leads to improved scatter on the photometric redshift. The scatter is high at low redshifts due to lack of $u$-band imaging. These results are consistent with those of \citet{Abdalla:DUNEphotoz}, \citet{DES:Whitepaper} and \citet{DES:photoz}. 

As can be seen in Figure \ref{fig:DESscatter} however, there are many outliers present in the sample. Another useful quantity to consider is therefore $\sigma_{68}$ which is the interval in which 68\% of the galaxies have the smallest difference between their spectroscopic and photometric redshifts. This will give us some indication of the scatter in the photometric redshift estimate once the outliers have been removed. We find that for DES $grizY$ photometry, $\sigma_z$ is 0.13 across the entire redshift range of $0<z<2$ whereas $\sigma_{68}$ is 0.08. When we add the VHS $JHK_s$ photometry to this, $\sigma_z$ improves to 0.11 across the entire redshift range and $\sigma_{68}$ improves to 0.07.

\subsection{Impact of Galactic Reddening}

\begin{figure}
\begin{center}
\includegraphics[width=8.5cm,angle=0]{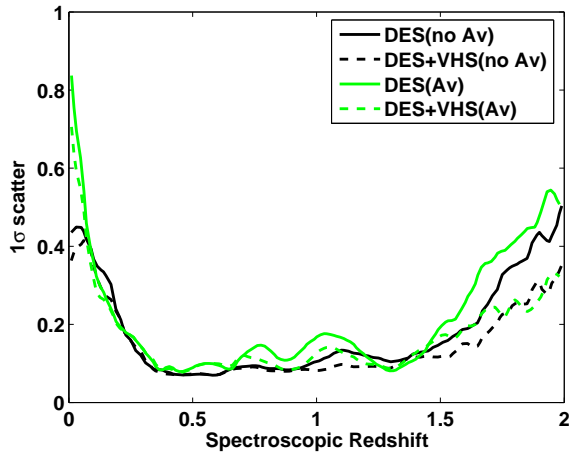}
\caption{The 1$\sigma$ scatter on the photometric redshift for DES with and without VHS NIR data for two different mock catalogues. The black lines are produced by DES5yr catalogues that do not include the effects of reddening. The green lines are produced by the JPLCAT mocks which include the effect of reddening. The solid lines show the scatter without VHS NIR data while the dashed lines include VHS NIR data. For both sets of mocks, the VHS NIR data improves the photo-z scatter by a factor of $\sim$2 at z$>1$ In regions of interest, the photoz scatter is worsened by $\sim$30\% when we include reddening in our mocks.}
\label{fig:DESsig2}
\end{center}
\end{figure}

In this section, we look at the impact of reddening on the photometric redshift estimate. We have already discussed how the DES5yr and JPLCAT samples differ in their inclusion of reddening in the galaxy samples. In order to assess how this difference affects the photo-z estimate, we run our neural network code on the JPLCAT sample with 5-band DES optical photometry as well as 8-band DES+VHS photometry with an exposure time of 120s in the NIR. The results are shown in  Figure \ref{fig:DESsig2} where we plot the $1\sigma$ scatter defined in Eq. \ref{eq:scatter} as a function of the spectroscopic redshift for both the DES5yr and JPLCAT samples for each of the two filter configurations. Note that before comparing the two catalogues, the JPLCAT sample has been cut to have the same magnitude and redshift limits as the DES5yr sample i.e. $0<z<2$ and $20<i<24$.

Although the same improvement is noted with inclusion of the NIR filters as discussed in $\S$\ref{sec:NIR}, we find that the effects of reddening worsen the photo-z scatter overall by $\sim30\%$ in some regions. This can be explained by the fact that there exists a degeneracy between redshift and galaxy reddening which means that faint reddened galaxies at low redshift can often appear to have the same colours as brighter galaxies at high redshift with no reddening \citep{Abdalla:DUNEphotoz}. However Figure 12 of \citet{Abdalla:DUNEphotoz} shows that this degeneracy is broken in the redshift range $1.1<z<1.5$ and we can see that the reddened DES only catalogues have a similar scatter to their unreddened counterparts in this redshift range. These authors have also shown that galaxies with small values of $A_v$ have relatively good photo-z estimates whereas those with high $A_v$ are scattered towards higher photometric redshifts. Figure \ref{fig:cats} shows that most of the galaxies in our JPLCAT sample have relatively small values of $A_v$ and therefore while we need to be aware that any amount of dust extinction is likely to affect our photo-z estimate, this effect should only be small for the DES sample.

In order to account for this effect of the dust extinction on the photometric redshift estimate, some authors attempt to include the dust extinction, $A_v$ as a free parameter in their codes (e.g. \citet{MRR:photoz, Bolzonella:HyperZ}) and simultaneoulsy solve for this and the photometric redshift. We have modified our neural network code to produce estimates for the $A_v$ and SED type of the galaxy using the JPLCAT sample with 8-band DES+VHS photometry. We use a 8:16:16:2 architecture for the neural network and marginalise over the redshift estimate. 

\begin{figure*}
\begin{center}
\begin{minipage}[c]{1.00\textwidth}
\centering 
\includegraphics[width=8.5cm,angle=0]{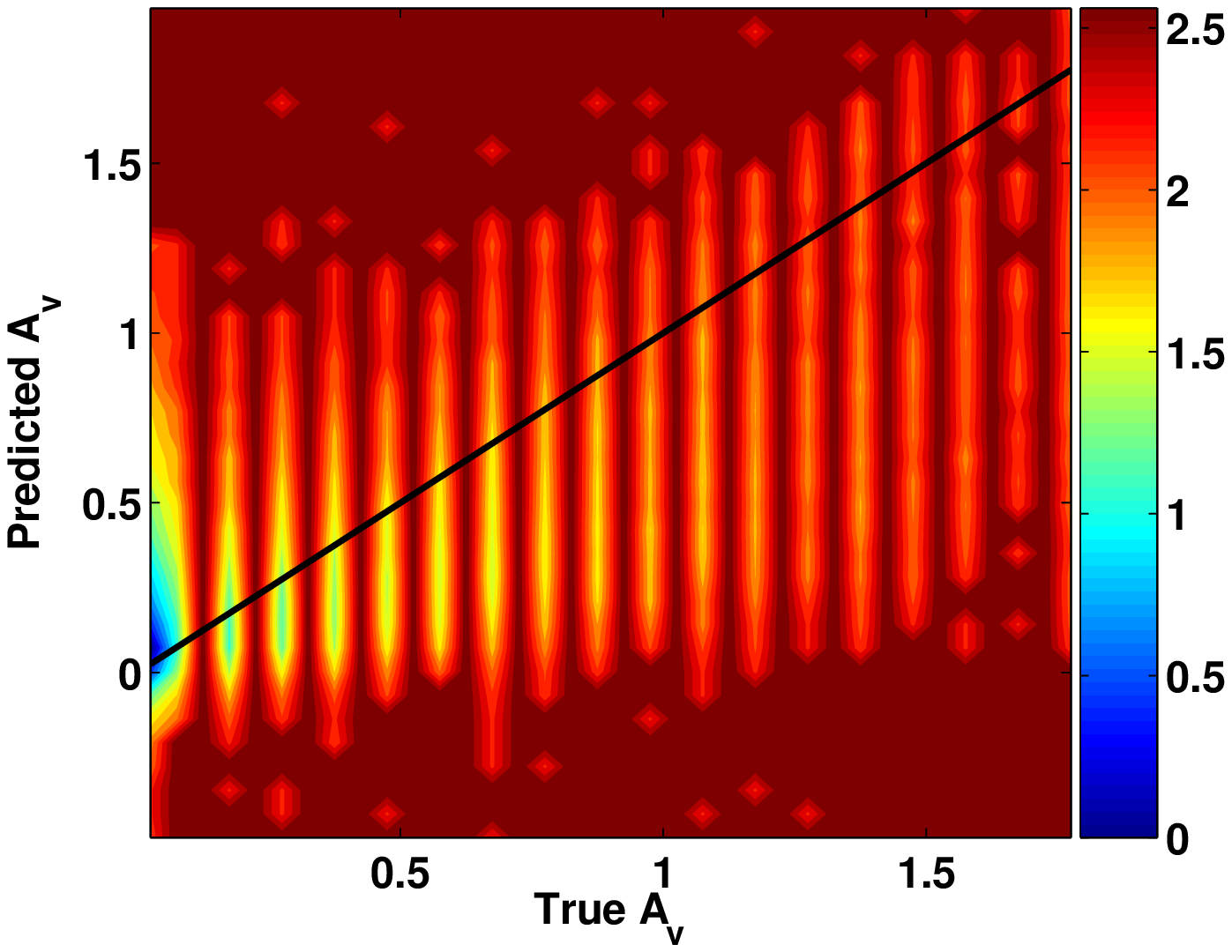}
\includegraphics[width=8.5cm,angle=0]{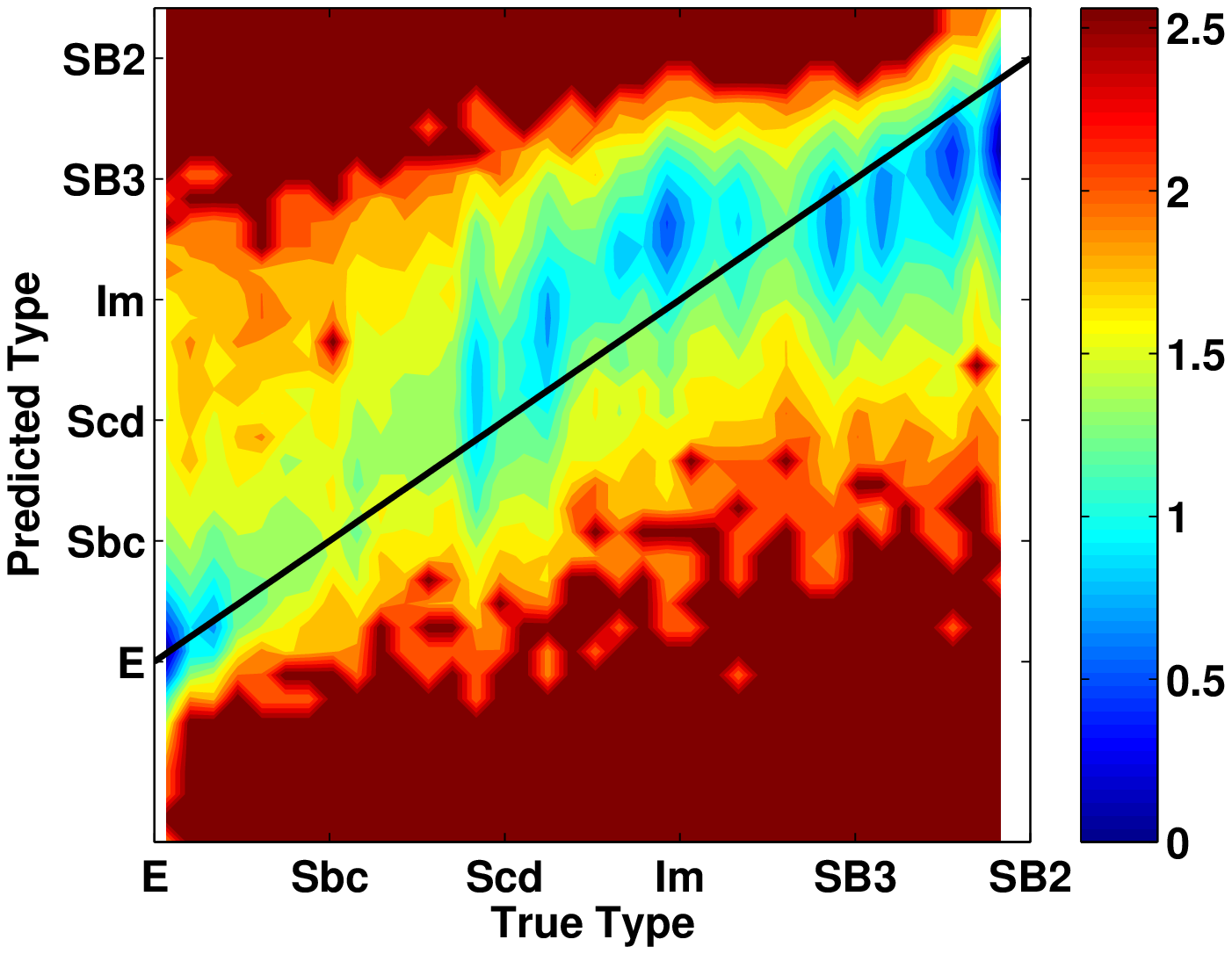}
\end{minipage}
\caption{Density plot of the ANNz output when the neural network code is used to simultaneously predict the $A_v$ and type of an object. The left-hand plot shows the predicted dust extinction, $A_v$ as a function of the true $A_v$. The right-hand plot shows the predicted SED type of each galaxy as a function of the true type. The plots are colour-coded and the scale is exponential; a colour difference of one is equivalent to the density being decreased by a factor of $e$. The solid black lines show where the true $A_v$ and true type are equal to the predicted $A_v$ and predicted type.}
\label{fig:Av}
\end{center}
\end{figure*}

The results are shown in Figure \ref{fig:Av} where we plot density plots of the true $A_v$ against the predicted $A_v$ and the true type against the predicted type.

We find the rms scatter around the mean of the $A_v$ estimate to be 0.27 and the bias to be 0.0031. The predicted $A_v$ is found to be biased towards lower values of $A_v$ for galaxies with a high degree of reddening and towards higher values of $A_v$ for galaxies with a low value of reddening. We also find the scatter on the type to be 7.7 and the bias to be -0.0048. The JPLCAT sample has been generated using six SED templates - E, Sbc, Scd, Im \citep{CWW:templates} and SB2, SB3 \citep{Kinney:temp} corresponding to types 0, 10, 20, 30, 40 and 50. As the error on the type is smaller than the difference between these templates, we can effectively use ANNz to classify our galaxies into $50/7.7 =$ about six or seven spectral types within the context of DES. 

Note that in this work, we have made no attempt to optimise our neural network for the calculation of the $A_v$ and type. We simply note that it is possible to use our neural network code to produce estimates for these quantities as well as the redshift and that this may be useful for samples where we know there is a high degree of reddening.  

Through the rest of this work, we have used the DES5yr sample for all the analysis. 

\subsection{Clipped Catalogues}

\label{sec:clip}

In the previous sections we have seen that catastrophic errors in the photometric redshift estimate can arise depending on the exact filter configuration and the galaxy formation science encoded within mock catalogues. Given that there are likely to be a host of different reasons why the photometric redshift estimate may be prone to large errors, a lot of which we do not fully understand, it seems sensible to devise some way of \textit{clipping} a sample. This is done by removing galaxies with large photo-z errors before using the photo-z estimate for cosmological analysis. In most situations where photometric redshift analysis is particularly powerful, we do not know the spectroscopic redshift of the galaxies and therefore have no way of using this information to assess whether the photo-z estimate is accurate. However, the photo-z prediction will depend strongly on the errors in the photometry and this information could potentially be used to \textit{clip} our sample as in \citet{Abdalla:DUNEphotoz}. In order to do this, we consider the neural network error estimate on the photometric redshift for each galaxy as given by Eq. \ref{eq:errANNz}. We then remove all galaxies from our sample that have an estimated error greater than a chosen threshold, therefore resulting in a \textit{clipped} catalogue of galaxies. We use the quantity, $\sigma_{68}$ introduced in $\S$ \ref{sec:NIR} to quantify the scatter in the photometric redshift estimate once the outliers have been removed. In this section, we extend the work of \citet{Abdalla:DUNEphotoz} and look at how the scatter on the photo-z estimate varies with different clipping thresholds.

\begin{figure*}
\begin{center}
\begin{minipage}[c]{1.00\textwidth}
\centering 
\includegraphics[width=8.5cm,angle=0]{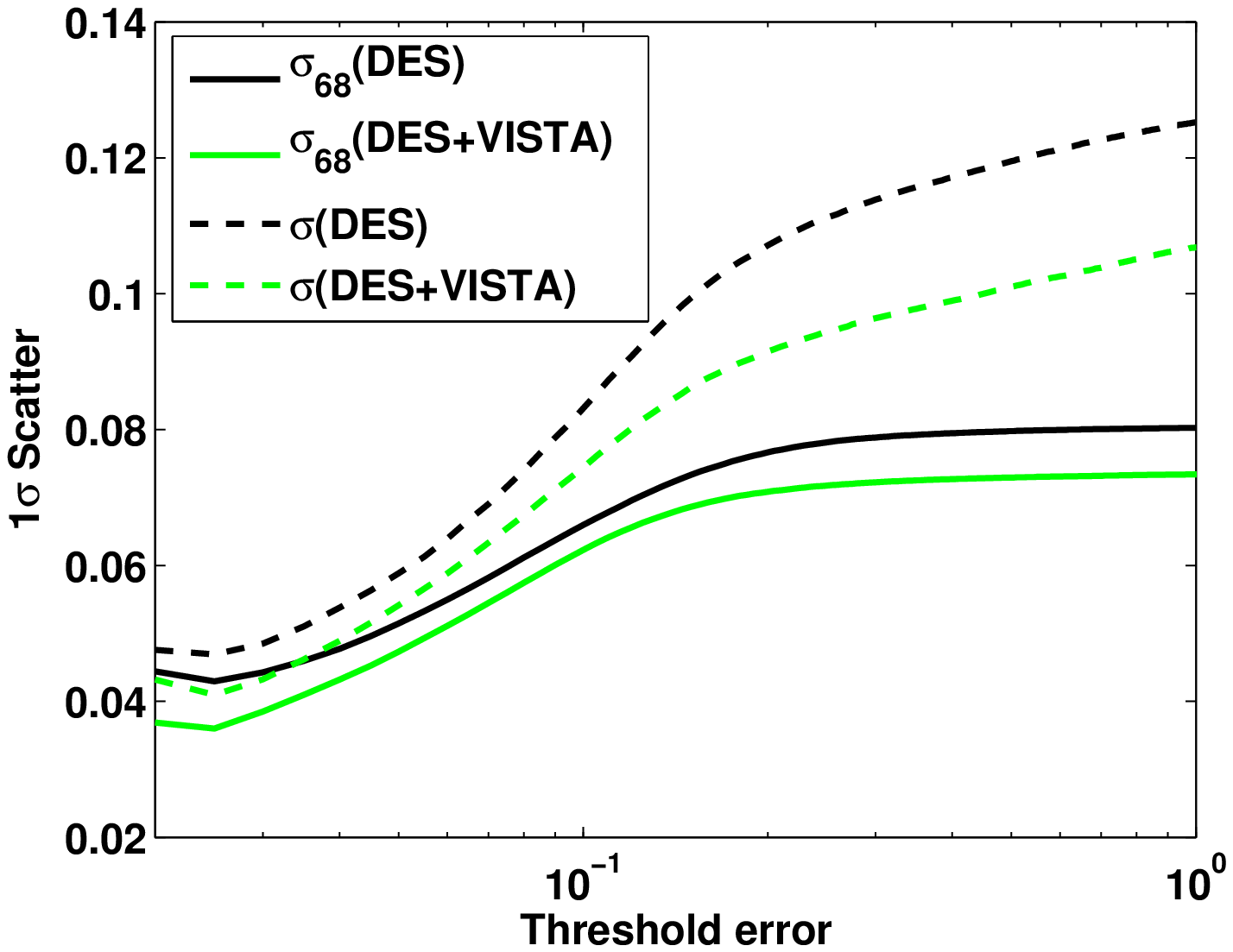}
\includegraphics[width=8.5cm,angle=0]{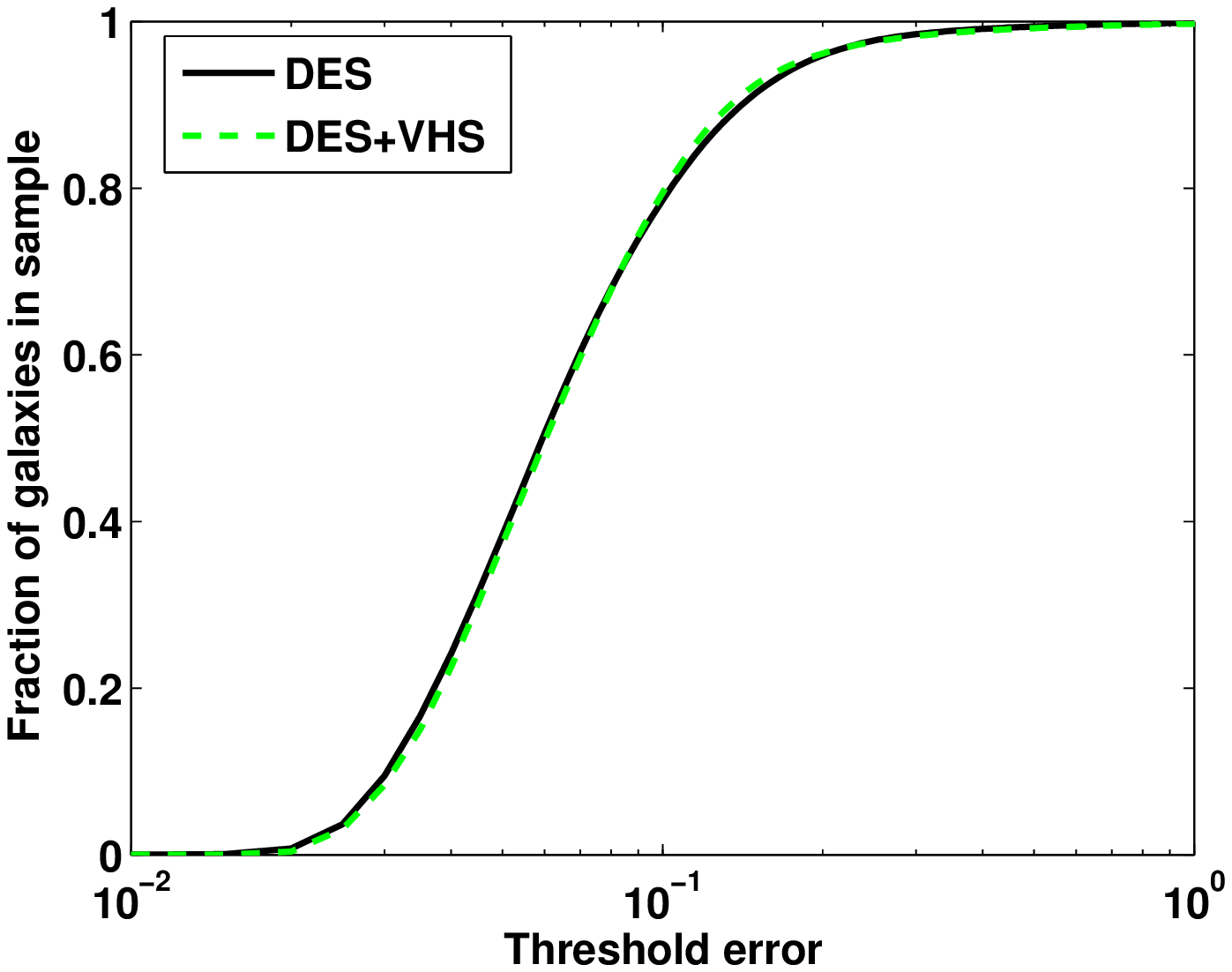}
\end{minipage}
\caption{The left hand plot shows the scatter, $\sigma_z$, and $\sigma_{68}$ as a function of the clipping threshold. The right-hand plot shows the fraction of galaxies remaining in the sample after the cuts are applied as a function of the clipping threshold. Galaxies with ANNz errors above the clipping threshold are removed from our sample.}
\label{fig:clipping}
\end{center}
\end{figure*}

Figure \ref{fig:clipping} illustrates the results of this study. In this figure we also plot the fraction of galaxies that remain in the sample once the clipping thresholds are applied. In each case, we plot results using a 5-band DES optical $grizY$ catalogue as well as an 8-band DES+VHS optical and NIR catalogue. 

As expected, applying smaller threshold errors at which to cut our sample results in a fall in the $1\sigma$ scatter for the entire sample. $\sigma_{68}$ also decreases as we reduce the threshold error although this decrease is less steep than the decrease in $\sigma_{z}$. Both the scatter and $\sigma_{68}$ are larger for the DES sample compared to the DES+VHS sample. We can see that applying a fairly conservative cut of 0.1 to our mock samples results in a reduction in $\sigma_{z}$ by a factor of $\sim1.5$ for both the DES and DES+VHS samples. In both cases we retain about 80\% of our original sample after this cut.We can apply smaller threshold errors in order to reduce the scatter on our photometric redshift estimate further. However, as we do this, we lose more galaxies from our original sample and at some point the number of galaxies remaining will prove insufficient for statistical analysis. We examine this point in more detail later in $\S$\ref{sec:cosmology}.

\subsection{Impact of Training Sets}

\subsubsection{Effect on the Photo-z Scatter}

All the photometric redshift analysis carried out in the previous sections assume that the training set used to train the neural network is totally representative of the testing set. However in reality, this may not always be the case. In this section, we look at the impact of using different training sets with different imposed colour and magnitude cuts on the photometric redshift estimate. 

The Dark Energy Survey  region overlaps with that of several other current and future photometric and spectroscopic surveys thereby providing it with a fairly complete sample of training set galaxies. Some of these are detailed in Table \ref{tab:training}. Here we consider two of the deeper surveys, namely DEEP2 and VVDS-Deep and model our training sets on the redshift distributions of these surveys before performing the usual photometric redshift analysis. Note that as DES overlaps the VVDS-Deep and DEEP2 fields, we assume that objects in these spectroscopic surveys will be imaged in all the DES bands. As the DES5yr sample is magnitude limited as described in $\S$\ref{sec:catalogues}, we do not consider the SDSS and 2dFGRS training sets as these objects are brighter than the mocks considered.

\begin{table*}
\begin{center}
    \begin{tabular}{|l|c|c|}
      \hline
        REDSHIFT SURVEY & SELECTION CRITERIA & NO.OF REDSHIFTS \\
	\hline
      SDSS Stripe 82 & $r<20$ & $70000$ \\
      2dFGRS & $b_{J}<19.45$ & $90000$ \\
      VVDS Shallow & $I_{AB}<22.5$ & $\sim 26000$ \\
      VVDS Deep & $I_{AB}<24$ & $\sim 60000$ \\
      DEEP2 & $(B-R)<0.4$ & $\sim 54000$ \\
            & $(R-I)>1.25$ & \\
            & $(B-R)<2.35(R-I)-0.54$ & \\
\hline
 \end{tabular}    \vspace{2mm}
  \end{center}
\caption{Summary of some of the spectroscopic surveys that will provide useful training sets for DES along with their imposed colour and magnitude cuts and the number of redshifts they are expected to obtain on completion.              \label{tab:training}} 
\end{table*}

DEEP2 is an ongoing spectroscopic survey being carried out by the DEIMOS spectrograph on the Keck II telescope. On completion, it will have obtained spectroscopic redshifts for $\sim54000$ objects over an area of 3.5deg$^2$. The survey has been designed to sample the redshift range of $0.75<z<1.5$ and the spectrograph is capable of obtaining moderately high resolution spectra between $6300 \AA$ and $9100 \AA$. Targets are pre-selected using $BRI$ imaging on the CFH12k camera on the Canada-France-Hawaii Telescope with a magnitude limit of $R_{AB}<24.1$ and the colour cuts detailed in Table \ref{tab:training} imposed in order to sample the redshift range of interest. 

\begin{figure}
\begin{center}
\centering
\includegraphics[width=8.5cm,angle=0]{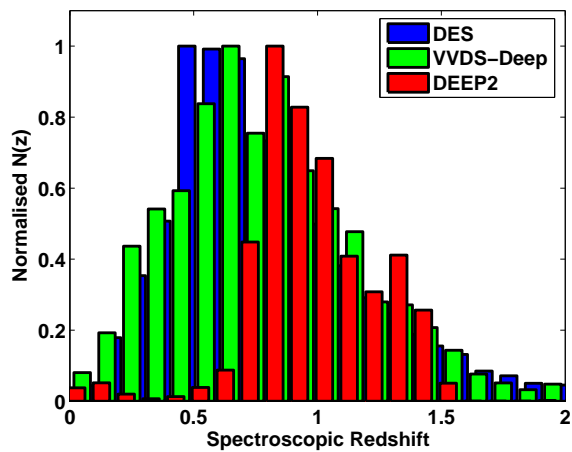}
\caption{The normalised redshift distributions for DES, VVDS-Deep and DEEP2 surveys between $0<z<2$. As can be seen, the DEEP2 colour cuts mean that most objects lie in the redshift range $0.7<z<1.4$. VVDS-Deep on the other hand effectively samples the entire DES redshift range although it has fewer galaxies at intermediate redshifts.}
\label{fig:dist}
\end{center}
\end{figure}

In this study, we use 4681 objects with spectra from DEEP2 DR1 \citep{Davis:DEEP2DR1} to construct the normalised redshift distribution for the DEEP2 survey. This is plotted in Figure \ref{fig:dist}. As can be seen, there are very few objects with redshifts less than $\sim0.7$ and greater than $\sim 1.4$. This is because the wavelength range for the spectrograph has been chosen such that the strong [OII] doublet which has a rest-frame wavelength of $3727 \AA$ lies outside these wavelengths for all other redshifts. Note that we have not included data from the Groth Survey Strip region in this study. This survey field has no imposed colour cuts and therefore may be useful for sampling the low-redshift range of DES.   

The VVDS spectroscopic surveys are being carried out using the VIMOS spectrograph on the Very Large Telescope (VLT). There is a shallow survey out to $I_{AB}=22.5$ planned in 5 fields and a deeper survey out to $I_{AB}=24$ in a single field. Targets are pre-selected using magnitudes from the imaging survey being carried out in the $UBVRI$ bands using the CFH12k camera on the CFHT.

The latest catalogue contains 8981 objects upto $I_{AB}=24$ in the redshift range $0<z<5.228$ \citep{LeFevre:VVDS1, LeFevre:VVDS2} and has been used to construct the normalised VVDS-Deep redshift distribution plotted in Figure \ref{fig:dist}. Note that for both the DEEP2 and VVDS-Deep samples, we have removed stars and other objects with very low redshifts as well as high-redshift objects with $z>2$ before plotting the redshift distributions so as to match the redshift range of the DES5yr sample. 

Having obtained the redshift distributions for both DEEP2 and VVDS-Deep, we proceed to construct accurate training sets that simulate these surveys to be used when running our neural network code. This is done as follows. We first seperate our DES5yr sample of 1 million objects into two equal sized training and testing sets. We then divide the training set into 20 redshift bins and from the DEEP2 and VVDS-Deep redshift distributions, calculate $N_i$, the number of galaxies from these surveys that would be present in each redshift bin, $i$ once the survey is complete and has obtained spectra for the number of objects given in Table \ref{tab:training}. We then randomly choose $N_i$ galaxies from the DES5yr training set to be put into redshift bin $i$ and in this way we construct a new training set of galaxies that have the same redshift distribution as our real spectroscopic surveys. The new training sets are then split further in order to create validation sets for ANNz to run on. This is done for a DES catalogue with optical $grizY$ photometry as well as a DES+VHS catalogue with 8-band optical and NIR $JHK_s$ photometry. Our simulated DEEP2 sample has many more galaxies at intermediate redshifts whereas the simulated VVDS-Deep survey samples the low and high redshift regimes better than DEEP2.

\begin{figure*}
\begin{center}
\begin{minipage}[c]{1.00\textwidth}
\centering 
\includegraphics[width=8.5cm,height=7.5cm,angle=0]{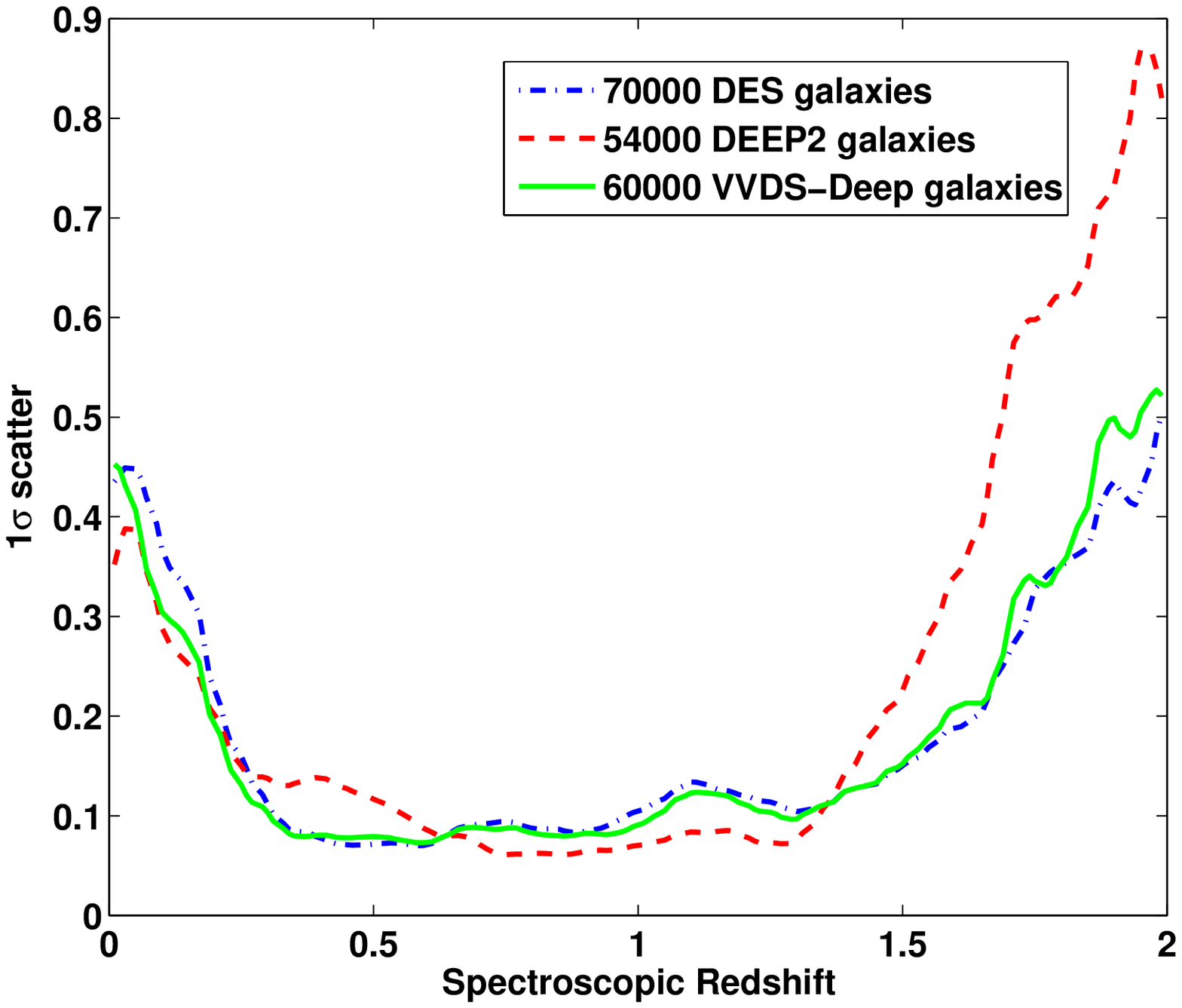}
\includegraphics[width=8.5cm,height=7.5cm,angle=0]{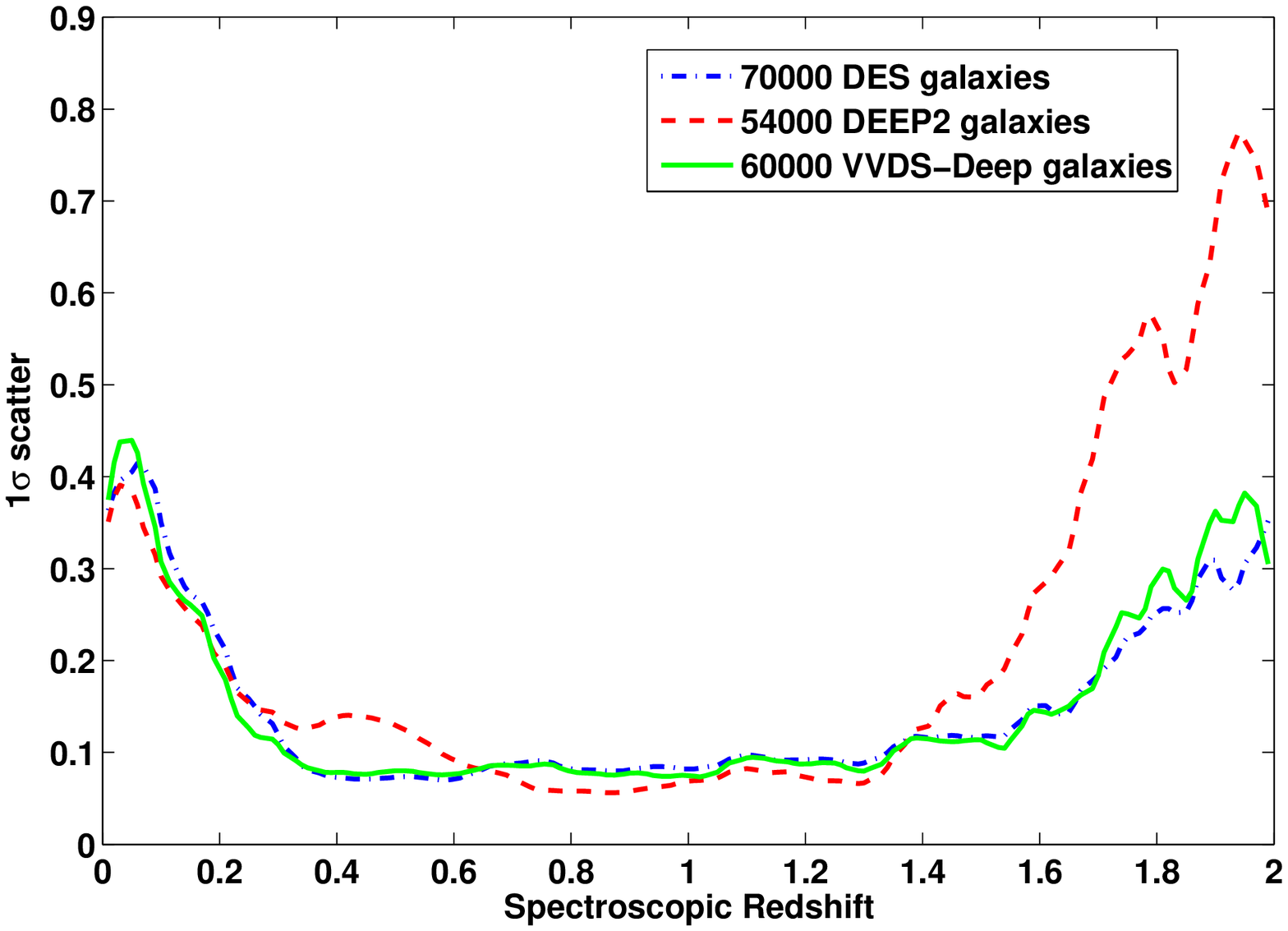}
\end{minipage}
\caption{The scatter on the photometric redshift as a function of the spectroscopic redshift when the DES, DEEP2 and VVDS-Deep redshift distributions are used to construct the training set used by the neural network. The left-hand plot shows the scatter for a catalogue with optical $grizY$ photometry and the right-hand panel shows the scatter for a catalogue with 8-band optical and NIR $JHK_s$ photometry. The scatter is big at low redshifts for all three training sets due to a lack of $u$-band photometry. At intermediate redshifts, the DEEP2 sample performs best as all its galaxies are concentrated in this redshift range. Both the DES and VVDS-Deep training sets produce considerably less scatter than the DEEP2 training set at high redshifts.}
\label{fig:n(z)}
\end{center}
\end{figure*}

We then run our neural network code on the two DES catalogues using three different training sets each time - a training set with a DES redshift distribution, one with a DEEP2 redshift distribution and one with a VVDS-Deep redshift distribution. The results are shown in Figure \ref{fig:n(z)} where we plot the scatter on the photometric redshift as a function of the spectroscopic redshift for all these cases.  

At low redshifts, the scatter is large for all three training sets due to lack of $u$-band data. At intermediate redshifts of $0.75<z<1.4$, the DEEP2 sample does better than the other training sets by $\sim$ 40\% as all its galaxies are concentrated in this region. As we move to even higher redshifts, the DEEP2 sample gives very poor results due to a lack of training set galaxies in this redshift range whereas the DES and VVDS-Deep samples perform better. For $1.4<z<2$, the DEEP2 scatter is worse by a factor of $\sim$2 compared to the VVDS-Deep and DES training sets. As expected, the scatter is smaller overall when we include NIR photometry for all three training set scenarios. The improvement is particularly noteworthy in the high redshift regime. Here, the scatter is reduced more for the VVDS-Deep and DES training sets with the addition of the NIR and not as much for the DEEP2 training set. 

We can therefore clearly see that using a combination of DEEP2 and VVDS-Deep data to calibrate our DES photometric redshifts, is already as good as having a complete training set for DES.   

\subsubsection{Effect on the Photo-z Bias}

\label{sec:bias}

The bias on the photometric redshift estimate, $b_z$, in a given redshift bin between $z_1$ and $z_2$ is given by:

\begin{equation}
b_{z}={\left<z_{spec}-z_{phot}\right>}
\label{eq:bias}
\end{equation}

This bias can arise from various sources. The performance of the neural network will introduce some difference between the photometric and spectroscopic redshifts. Furthermore, having an incomplete training set or a cosmic variance limited sample also leads to biases in the photometric redshift estimate. If this bias does not depend on the testing set however, we can quantify it exactly using our training set and it can be subtracted from the photometric redshift estimate no matter how large it is. Once this is done, the residuals give us some indication of the dependance of the bias on the testing set. This error on the bias cannot be corrected for and it is this quantity that we evaluate in this section.  

We quantify the errors in the bias that arise from using different training and testing sets. In particular we look at the effects of size and incompleteness of both the training and testing sets. All the analysis carried out here is for the DES+VHS dataset and we model the incomplete training sets by imposing the DEEP2 colour cuts detailed in Table \ref{tab:training} on galaxies from the DES5yr catalogue. 

The standard deviation on the bias in each bin can be defined as follows assuming Poisson statistics \footnote{We have checked that this is a reasonable approximation to the error arising from the neural network and the analytical expression for the error on the bias agrees with the error from the neural network to within 15\%.}

\begin{equation}
rms(b_z)=\frac{\sigma_z}{\sqrt{N_s}}
\label{eq:bias}
\end{equation}

\noindent where $\sigma_z$ is the $1\sigma$ scatter on the photometric redshift given by Eq. \ref{eq:scatter} and $N_s$ is the number of spectroscopic training set galaxies. Given a suitably large number of training set galaxies with spectroscopic redshifts, this error on the bias can be effectively ignored as it does not depend on the testing set. In order to better understand some of the other sources of error on the bias that do depend on the testing set, we study the samples detailed in Table \ref{tab:sets}. 

\begin{table}
  \begin{center}
    \begin{tabular}{|l|c|c|}
      \hline
       & Number of Galaxies & Colour Cuts \\
      \hline
      SET1 & 50000 & None \\
      SET2 & 50000 & Same as DEEP2 \\
      SET3 & 70000 & None \\
      SET4 & 200000 & None \\
      SETa & 250000 & None \\
      SETb & 250000 & None \\
      SETc & 102643 & Same as DEEP2 \\
      SETd & 102589 & Same as DEEP2 \\
      SETe & 500000 & None \\
      \hline
    \end{tabular}     \vspace{2mm}
    \end{center}
  \caption{Summary of training and testing sets used to quantify differences in estimates of the photometric redshift bias. \label{tab:sets}}
\end{table}

\begin{figure*}
\begin{center}
\begin{minipage}[c]{1.00\textwidth}
\centering 
\includegraphics[width=8.5cm,height=7.5cm,angle=0]{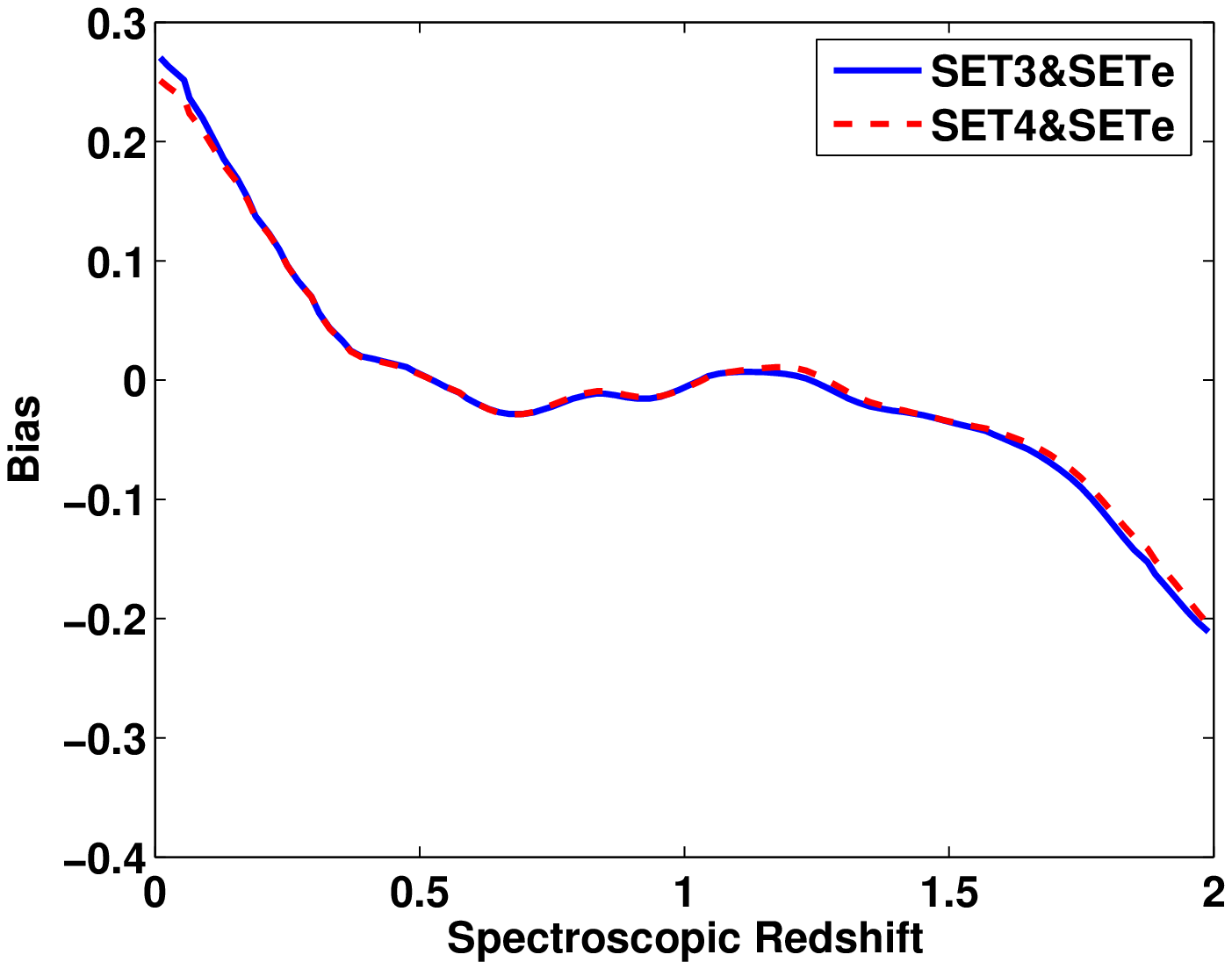}
\includegraphics[width=8.5cm,height=7.5cm,angle=0]{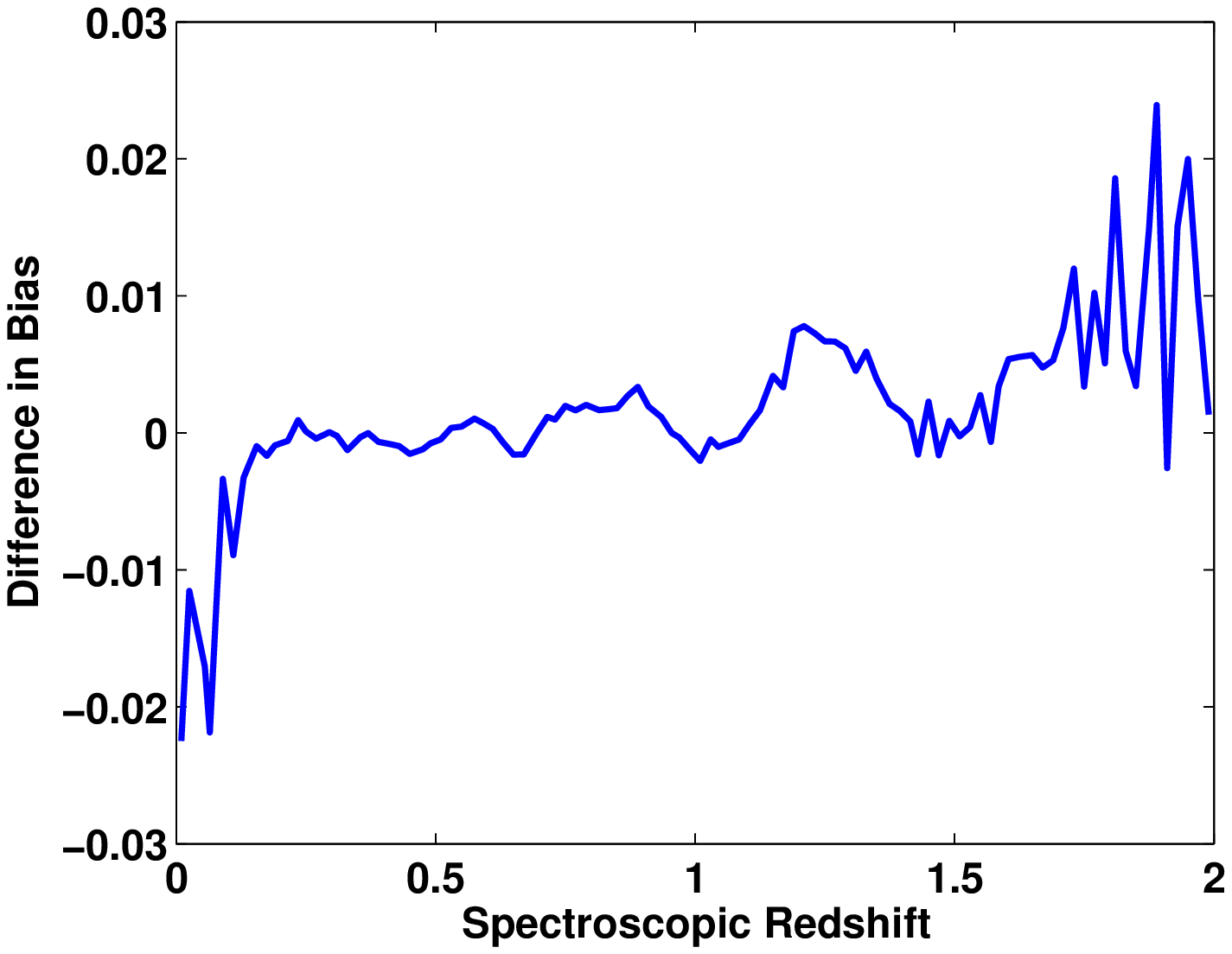}
\end{minipage}
\caption{The bias on the photometric redshift estimate when using a training set of 70000 galaxies (SET3) and when using a training set of 200000 different galaxies (SET4) on the same testing set. The right hand panel plots the difference between the two biases. We can see that increasing the number of training set galaxies by a factor of $\sim3$ leads to a change in the bias of the order of $10^{-3}$.}
\label{fig:nv}
\end{center}
\end{figure*}

We first quantify the difference in the bias when using different numbers of training set galaxies to calculate photometric redshifts for the same sample of testing set galaxies. Note that the differences in the bias calculated here are used as empirical estimators of the systematic shift one would get when dealing with data. SET3 and SET4 are used as training sets to calculate photometric redshifts for SETe. All three samples are complete and have no imposed colour cuts. In Figure \ref{fig:nv} we plot the biases obtained for each of the two cases and the difference between these. This shows us that changing the size of our training set by a factor of $\sim3$ leads to a difference in the biases of the order of $10^{-3}$.

\begin{figure*}
\begin{center}
\begin{minipage}[c]{1.00\textwidth}
\centering 
\includegraphics[width=8.5cm,height=7.5cm,angle=0]{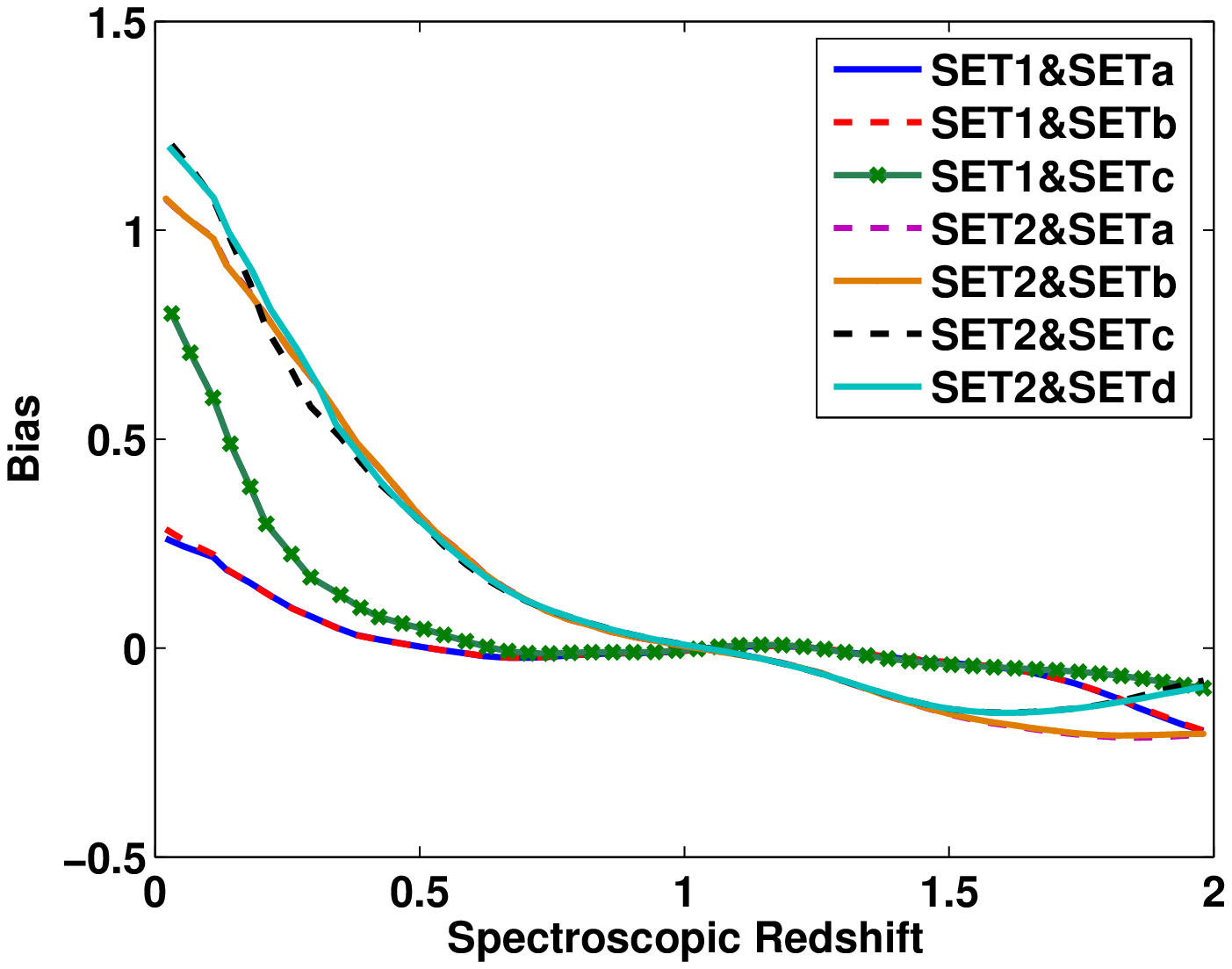}
\includegraphics[width=8.5cm,height=7.5cm,angle=0]{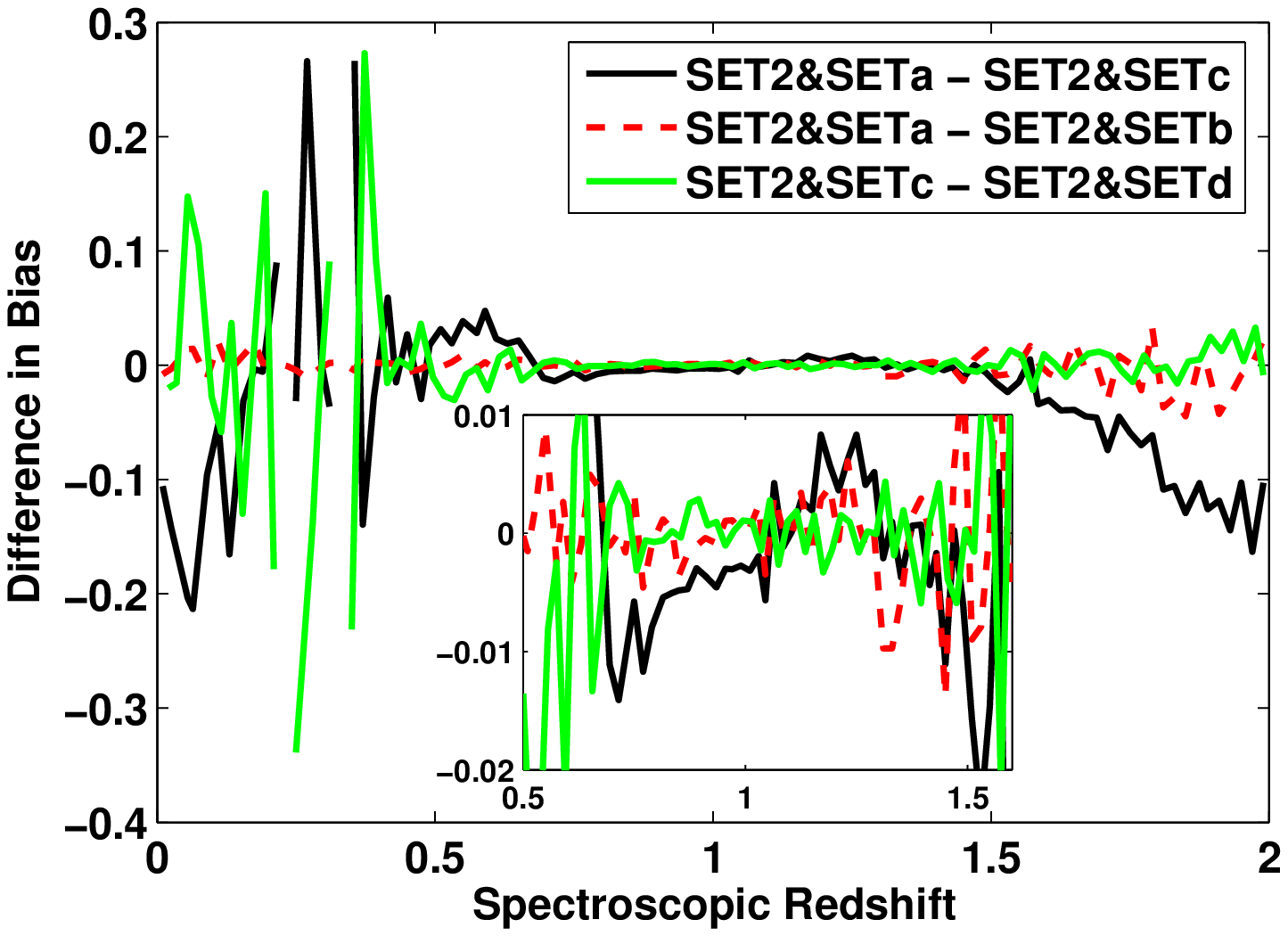}
\end{minipage}
\caption{The left-hand panel shows the biases on the photometric redshift estimate when using each of the different configurations of training and testing sets detailed in Table \ref{tab:sets}. In the right-hand panel, we plot the difference in the bias when using two different sets of DES5yr galaxies with an incomplete training set (broken line), two different sets of DEEP2 galaxies with an incomplete training set (green solid line) and the difference in the bias when using a complete testing set and a testing set with the same colour cuts as the incomplete training set (black solid line). These plots are generated using bins of width 0.04 in redshift space.}
\label{fig:biases}
\end{center}
\end{figure*}

We proceed now to look at effects of incomplete training sets on the photometric redshift bias. To do this, we use SET1 and SET2 as training sets to calculate the photometric redshifts for samples SETa, SETb, SETc and SETd. SET1 is a complete training set while SET2 has been cut to reflect the colour cuts of the DEEP2 survey. SETa and SETb are both complete testing sets with different galaxies in them from the DES5yr mock catalogue, while SETc is generated by imposing the DEEP2 colour cuts on SETb and SETd by imposing the DEEP2 colour cuts on SETa. The biases on the photo-z estimate obtained for each of the different configurations of training and testing sets, are shown in Figure \ref{fig:biases}. Note that throughout this analysis, we use bins of width 0.04 in redshift space. 

From Figure \ref{fig:biases} we can draw the following general conclusions. Changing the galaxies that are present in the testing set when training with a complete and fully representative training set, leads to a change in the bias of the order $10^{-3}$. Using an incomplete training set on these samples also leads to the same difference in the bias between them (Figure \ref{fig:biases} - right panel, broken line). When using an incomplete training set such as that provided by the DEEP2 survey (SET2) on a testing set of galaxies with no imposed colour cuts (SETa and SETb), the bias in the photometric redshift is worsened as expected. However, if we impose the same colour cuts as DEEP2 on our testing set (SETc), the bias is improved by $\sim$20\% for $z<0.7$ and by $\sim$40\% for $z>1.4$ - i.e. in the redshift ranges where the training set is incomplete. The difference in the bias when using a testing set with no colour cuts and one with imposed colour cuts to match those of the incomplete training set, is of the order of $5\times 10^{-3}$ and always smaller than $10^{-2}$ in the redshift range of the incomplete training set - inset, left-hand panel of Figure \ref{fig:biases}. 

In $\S$\ref{sec:cosmology2}, we will briefly comment on how the errors on the photometric redshift bias propagate into errors in the calculation of the dark energy equation of state parameter, $w$.  

\section{Implications for Cosmology}

\subsection{Optimal Estimation of the Galaxy Power Spectrum}

\label{sec:cosmology}

In this section, we look at the impact of photometric redshift estimation on dark energy science. In particular, we concentrate on the measurement of dark energy using galaxy power spectra and baryon acoustic oscillations. 

The galaxy angular power spectrum is a measure of the clustering in the galaxy population within time bins extending from the present to a time when the Universe was only a third of its present age. Large-scale surveys like DES provide ideal data sets for studying the clustering properties of galaxies and therefore the clustering properties of their underlying dark matter distribution and hence are useful probes for mapping how the dark matter distribution evolves with time. Furthermore, many other characteristic features appear in the power spectrum which provide \textit{standard rulers} that can be used to determine the angular diameter distance, $D_A$, as a function of redshift. Baryon acoustic oscillations are one such feature of interest which appear as \textit{wiggles} on the power spectrum. The position of the peaks and troughs of these wiggles in Fourier space can be used to determine a set of cosmological parameters e.g. \citet{Blake:03} and \citet{Seo:03}. 

The accuracy with which we can measure this typical acoustic scale is proportional to the average fractional error in the power spectrum, $\delta P/P$. The fractional error on the power spectrum arises from two sources. Firstly, the number of independant spatial modes that we can measure in a given volume is finite and this will lead to errors in the power spectrum that are proportional to $1/\sqrt{V}$. This is known as cosmic variance. Secondly, there is a contribution from shot noise due to imperfect sampling of the fluctuations as we only have a finite number of tracers of these fluctuations within a given volume. If we assume a density field that follows Gaussian statistics, we can follow \citet{FKP:PowerSpec} and assume the error on the power spectrum measurement, $P$ is weighted in the following way:

\begin{equation}
\frac{\delta P}{P}\propto \frac{1}{\sqrt{V}}\left(1+\frac{1}{nP}\right)
\label{eq:errorps}
\end{equation}

\noindent where $n$ is the mean number density in a given volume as seen by an observer and can be written in terms of the galaxy redshift distribution as follows:

\begin{equation} 
\frac{dN}{dz}=f_{sky}\frac{dV}{dz}n(z)
\label{eq:n(z)}
\end{equation}

\noindent where $f_{sky}$ is the fraction of the sky covered by the survey and $dV/dz$ is the comoving volume element.

The first term in Eq. \ref{eq:errorps} denotes the effect of cosmic variance while the second term is the contribution from shot noise. In order to minimise the error on the power spectrum, one has to design a survey with maximum volume provided there are enough sources within this volume for the shot noise contribution to be minimal. If $nP>3$ the power spectrum is well estimated and there is no significant advantage to be gained with more galaxies \citep{Seo:03}. In this work, we assume that to obtain a reasonable estimate of the power spectrum we need to satisfy the condition, $nP \gtrsim 1$. Taking into account the galaxy bias, $b$ that scales the galaxy power spectrum to the matter power spectrum, and including the scaling of the matter power spectrum with redshift as a linear growth factor, $D(z)$ we get the following expression for $nP_{gal}$:

\begin{equation}
n(z)P_{gal}(k_*)=n(z)b^2(z)D^2(z)P(k_*)
\label{eq:ps2} 
\end{equation}

We have used the formalism for the transfer function set out in \citet{EH:transfer} to calculate our power spectrum at $k_*=0.1h{Mpc}^{-1}$ as this is well within the linear regime of the power spectrum. At larger values of $k$, non-linearities due to clustering and other structure formation start to dominate and make it harder to detect the BAO signal.

\begin{figure*}
\begin{center}
\includegraphics[width=8.5cm,height=7.5cm,angle=0]{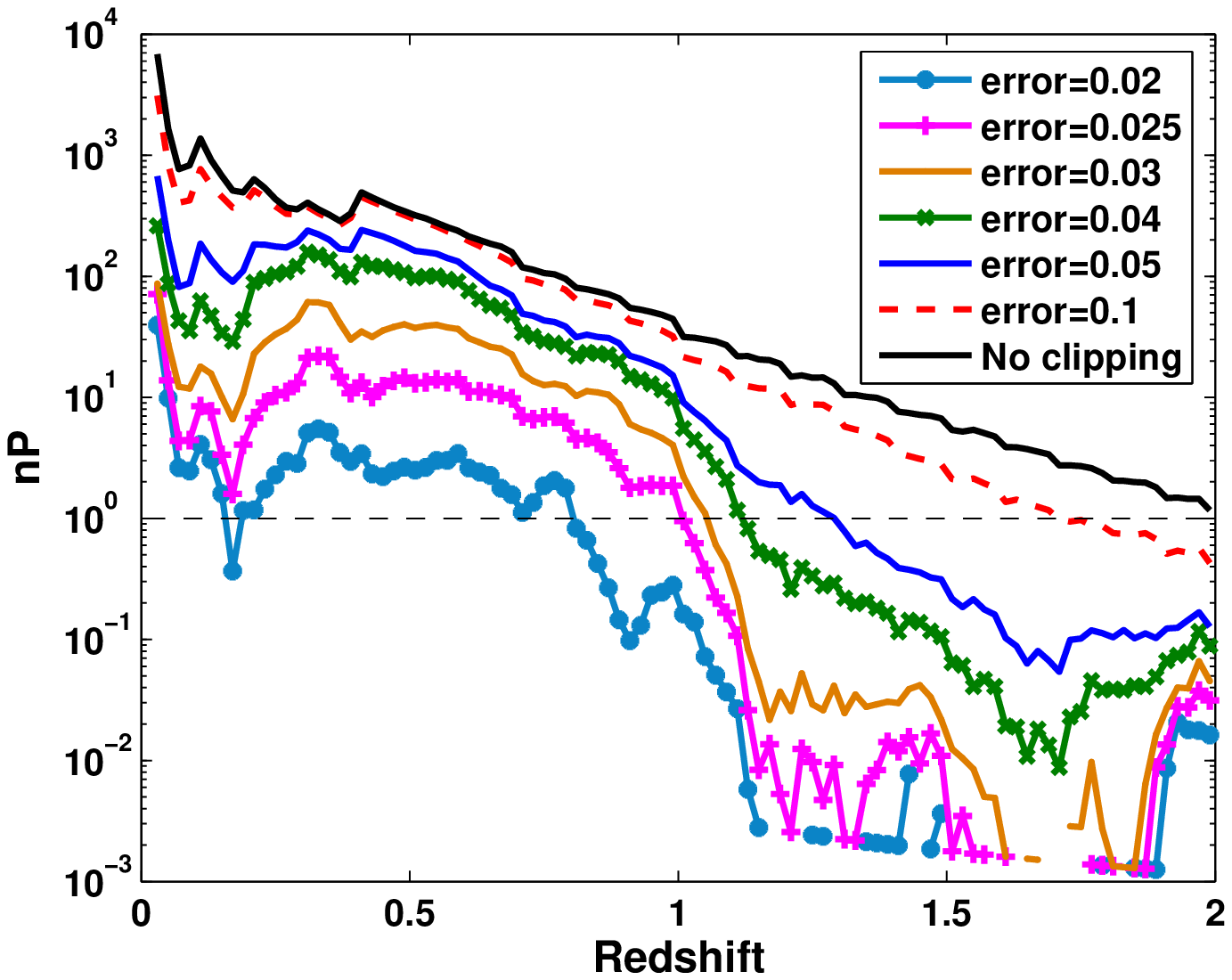}
\includegraphics[width=8.5cm,height=7.5cm,angle=0]{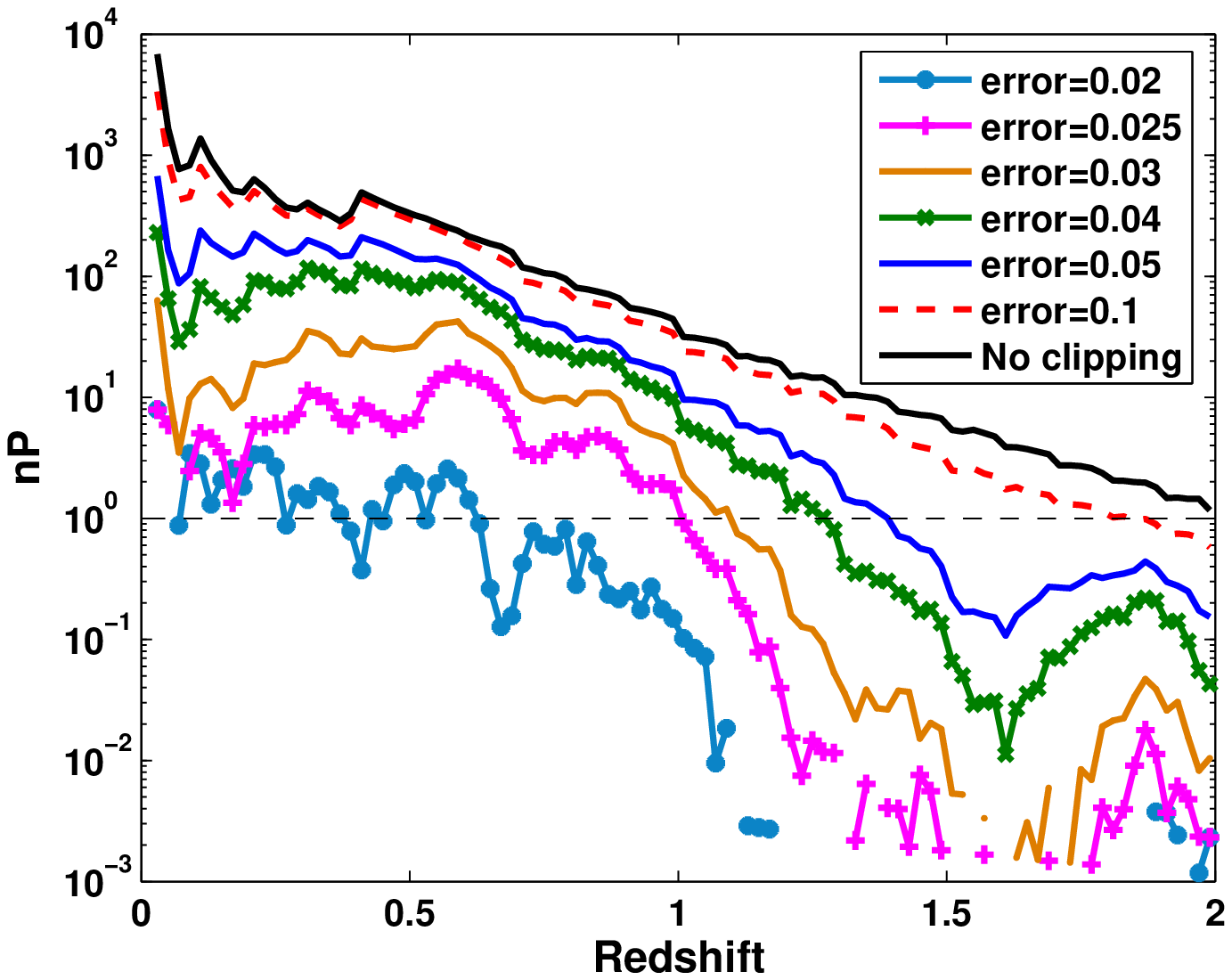}
\caption{$nP$ as a function of the redshift for different levels of clipping. The left-hand plot is generated using DES $grizY$ photometry only whereas the right-hand plot is produced from a catalogue with 8-band DES+VHS photometry. For the same clipping threshold error, the DES+VHS power spectrum is greater than 1 over a larger redshift range compared to the DES only case. This means that we can obtain an optimal measurement of the power spectrum out to higher redshifts if we use the full 8-band DES + VHS photometry.}
\label{fig:PowerSpec}
\end{center}
\end{figure*}

We assume a survey with $0<z<2$ and $f_{sky}=0.119$. The bias is assumed to be 1.2 \footnote{While we are aware of the dependance of our results on this bias, it is difficult at this point to make an educated guess of what $b$ will be for DES galaxies. We have therefore used a reasonable scale independant bias in our calculations.}. In Figure \ref{fig:PowerSpec}, we plot $nP_{gal}$ as a function of the redshift. This is done for the entire catalogue and for clipped catalogues with different clipping thresholds. We perform the same analysis for optical only DES data as well as optical and NIR data from DES+VHS. The results are summarised in Table \ref{tab:powerspec} and Figure \ref{fig:PowerSpec}.

\begin{table*}
\begin{center}    
\begin{tabular}{|l|c|c|c|c|}
      \multicolumn{5}{c}{\textbf{DES $grizY$ photometry}} \\
      \hline
      Threshold Error & Redshift Range for $nP\gtrsim1$ & $\sigma$ & $\sigma_{68}$ & Fraction of Galaxies Remaining \\
      \hline
      None & $0<z<2$ & 0.128 & 0.08 & 1 \\
      0.100 & $0<z<1.7$ & 0.084 &0.065 & 0.79 \\
      0.050 & $0<z<1.3$ & 0.058 & 0.052 & 0.38 \\
      0.040 & $0<z<1.1$ & 0.055 & 0.047 & 0.24 \\
      0.030 & $0<z<1.05$ & 0.048 & 0.044 & 0.095 \\
      0.025 & $0<z<0.95$  & 0.047 & 0.043 & 0.035 \\
      0.020 & $0<z<0.8$ & 0.047 & 0.045 & 0.007 \\
    \hline
\end{tabular}    \vspace{2mm}
  \end{center}

\begin{center}
    \begin{tabular}{|l|c|c|c|c|}
      \multicolumn{5}{c}{\textbf{DES $grizY$ + VHS $JHK_s$ photometry}} \\ 
      \hline
Threshold Error & Redshift Range for $nP\gtrsim1$ & $\sigma$ & $\sigma_{68}$ & Fraction of Galaxies Remaining \\
    \hline
    None & $0<z<2$  & 0.11 & 0.074 & 1 \\
      0.100 & $0<z<1.9$ & 0.074 &0.062 & 0.80 \\
      0.050 & $0<z<1.4$ & 0.054 & 0.048 & 0.37 \\
      0.040 & $0<z<1.3$ & 0.049 & 0.043 & 0.22 \\
      0.030 & $0<z<1.1$  & 0.043 & 0.039 & 0.09 \\
      0.025 & $0<z<1.0$  & 0.041 & 0.037 & 0.03 \\
      0.020 & $0<z<0.65$ & 0.041 & 0.037 & 0.005 \\
\hline
 \end{tabular}    \vspace{2mm}
  \end{center}

\caption{Summary of the redshift ranges over which we can obtain optimal measurements of the power spectrum for different clipping threshold errors and the corresponding values of $\sigma$, $\sigma_{68}$ and the fraction of galaxies remaining in our sample for each of these cases. The top table is for DES $grizY$ photometry and the bottom table for DES+VHS $JHK_s$ photometry. \label{tab:powerspec}} 
\end{table*}

From these results we can see that applying a threshold error at which to cut our photometric redshift catalogue proves effective in removing outliers from our sample before performing any kind of cosmological analysis on it. For the DES catalogue of redshifts obtained using $grizY$ photometry, we can remove all galaxies with a threshold error of more than $\sim0.03$ in order to obtain an accurate measurement of the galaxy power spectrum out to a redshift of 1. This leaves us with only 10\% of our original sample but this sample has a scatter on its photometric redshift that is a factor of $\sim2.7$ times better than that of the original sample and is therefore more effective in constraining the cosmology. For the DES + VHS catalogue, a threshold error of 0.025 can be applied to effectively constrain the galaxy power spectrum to redshift 1. This leaves us with only 3\% of our original sample, but the overall scatter on the photometric redshift has been reduced by a factor of $\sim2.75$. This is equivalent to performing an LRG selection on our survey as these galaxies have more accurate large-scale structure signals and more accurate photometric redshifts due to the prominence of their 4000\AA\@ break \citep{Blake:CosmoLRG,Pad:LRGphotoz}. 

In order to provide a reasonable measurement of the galaxy power spectrum for the entire DES redshift range of $0<z<2$, we can apply a threshold error cut of $>0.1$ to the DES only sample and use most of the galaxies in our analysis. When we add NIR photometry from VHS to our sample, a less conservative clipping cut of 0.05 can be applied and only 37\% of the galaxies used to reduce the scatter on the photometric redshift by a factor of $\sim2$. Note that $\sigma_{68}$ is also reduced in these cases although not to the same extent as the reduction in $\sigma$. A reduction in $\sigma_{68}$ corresponds to a reduction in the intrinsic scatter of our sample minus the outliers.

By clipping our catalogues in this way before performing any kind of cosmological analysis on them, we have effectively managed to reduce the errors in our measurement of the galaxy power spectrum without compromising on the precision with which this measurement has been made. Adding NIR data from VHS to our DES photometry has also allowed us to clip our catalogues more effectively and therefore make more precise measurements of the galaxy power spectrum out to higher redshifts. 

We now follow \citet{Blake:BAOfit} and \citet{BlakeBridle:Cosmo} and take into account the photometric redshift errors explicitly in our galaxy power spectrum analysis. These authors have shown that the fractional error on the galaxy power spectrum is related to the photometric redshift error as follows:


\begin{equation}
\frac{\delta P}{P}\propto\sqrt{\sigma_r}
\label{eq:photozps}
\end{equation}

\noindent where $\sigma_r$ is the rms error in comoving coordinates in units of $h^{-1}Mpc$. We can relate this to the redshift error already introduced in Eq. \ref{eq:scatter} as follows:

\begin{equation}
\sigma_r=\sigma_z\frac{c}{H(z)}=\frac{\sigma_zc}{H_0\sqrt{(\Omega_m(1+z)^3+\Omega_k(1+z)^2+\Omega_{\Lambda})}}
\label{eq:photozcom}
\end{equation}

The fractional error in the power spectrum is now given by:

\begin{equation}
\frac{\delta P}{P} \propto \sqrt{\frac{\sigma_r}{V}}\left(1+\frac{1}{nP}\right)
\label{eq:errorps2}
\end{equation} 

We can calculate this quantity for different values of the clipping threshold. This is done for an optical DES sample as well as an optical and NIR DES+VHS sample in redshift bins of width 0.02. We assume a DES comoving survey volume, V of 23.7$h^{-3}Gpc^3$ between $0<z<2$. Our choice of redshift is motivated by Figure \ref{fig:PowerSpec} where it can be seen that there are enough galaxies in the survey out to a redshift of 2 for shot-noise errors not to be significant, provided we do not clip our sample. As we are calculating the fractional error in the power spectrum in redshift bins, we define the effective volume in a redshift bin as follows:

\begin{equation}
\delta V_{eff}=\left[\frac{n(z)P}{(n(z)P+1)}\right]^2 \delta V
\label{eq:Veff}
\end{equation}

The fractional error on the power spectrum in a redshift bin can then be re-written as:

\begin{equation}
\frac{\delta P}{P} \propto \sqrt{\frac{\sigma_r}{\delta V_{eff}}}
\label{eq:errorps3}
\end{equation}

As we can see, decreasing the threshold error at which to clip the sample reduces the scatter on the photometric redshift and therefore $\sigma_r$ leading to smaller fractional errors on the power spectrum. However this also reduces $nP$ and therefore $\delta V_{eff}$, thereby increasing the shot-noise contribution to the error in the power spectrum due to a lack of sufficient galaxies in the sample. Clearly, there is a threshold error that needs to be determined and this is what we proceed to do.  

\begin{figure*}
\begin{center}
\includegraphics[width=8.5cm,height=7.5cm,angle=0]{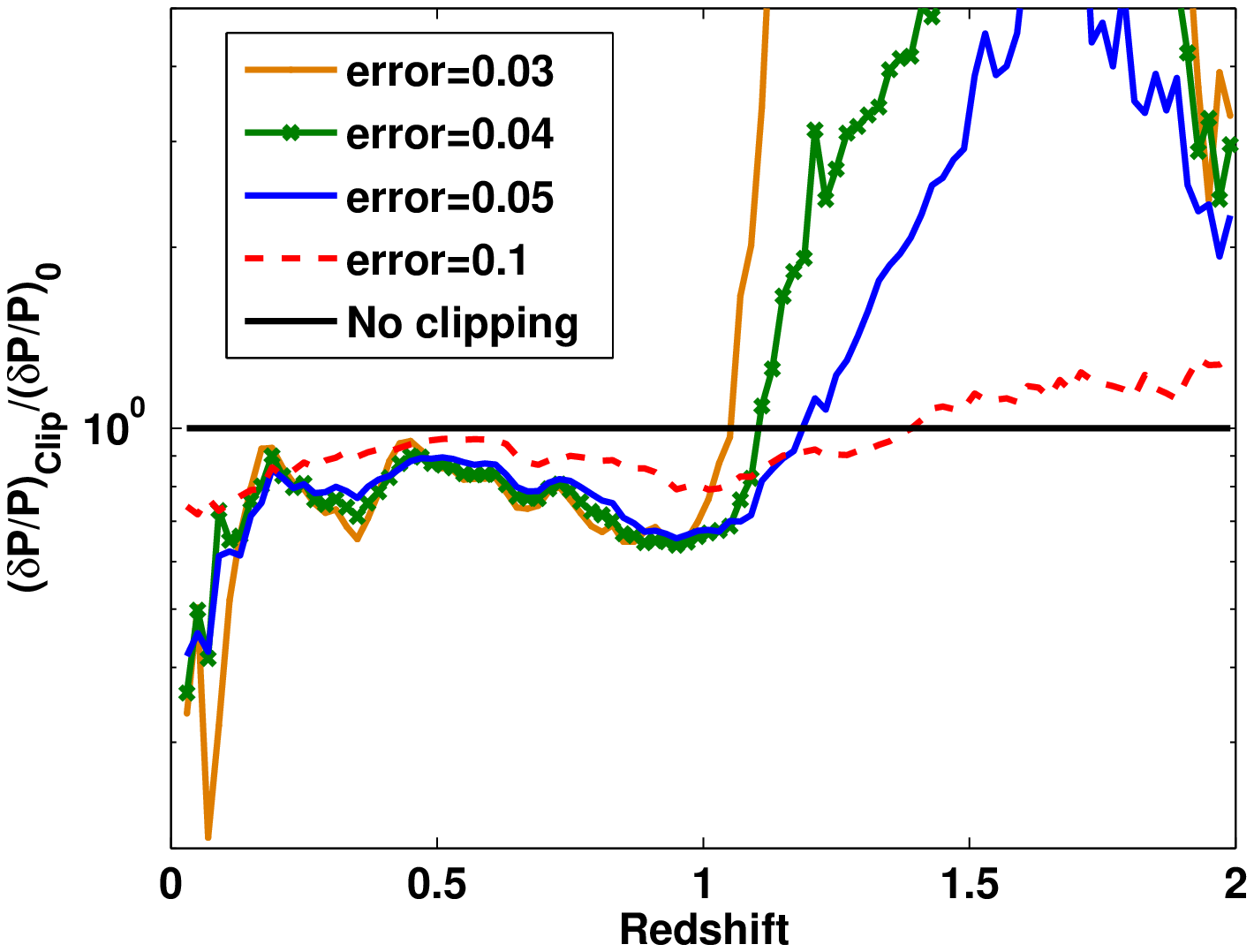}
\includegraphics[width=8.5cm,height=7.5cm,angle=0]{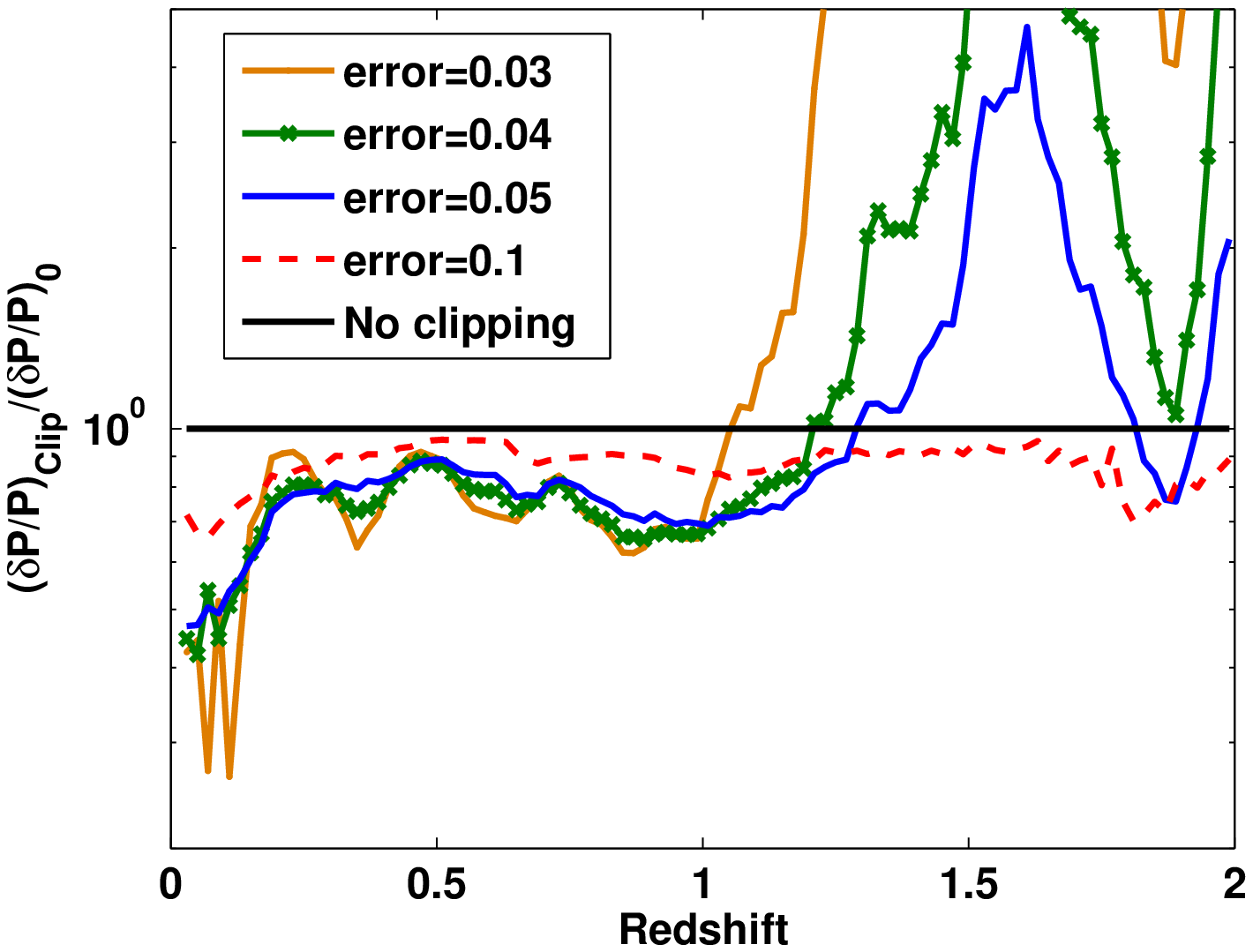}
\caption{$\delta P/P$ for different levels of clipping divided by  $\delta P/P$ for no clipping as a function of the redshift. The left-hand plot is generated using DES $grizY$ photometry only whereas the right-hand plot is produced from a catalogue with 8-band DES+VHS photometry. At small threshold errors, the power spectrum measurement is shot-noise dominated whereas at large threshold errors, large photo-z errors result in large positional uncertainties.}
\label{fig:deltaP}
\end{center}
\end{figure*}

The results of our study are shown in Figure \ref{fig:deltaP} and Table \ref{tab:deltaP} where we plot the fractional error in the power spectrum for different levels of clipping divided by the fractional error in the power spectrum obtained using the entire sample. We only do this for clipping thresholds greater than or equal to 0.03 as below this, shot noise is dominant across the entire redshift range. If the plotted quantity is less than one for a given clipping threshold, cutting the sample using this threshold improves our constraints on the galaxy power spectrum.  

\begin{table*}
\begin{center}    
\begin{tabular}{|l|c|c|c|}
      \multicolumn{4}{c}{\textbf{DES $grizY$ photometry}} \\
      \hline
  Redshift Range & Optimum Clipping Threshold & Fraction of galaxies used & Improvement in $\delta P/P$ \\
      \hline   
      $0<z<0.1$ & 0.025 & 4\% & 75\%\\
      $0.1<z<0.9$ & 0.03 & 10\% & 20-30\% \\
      $0.9<z<1.0$ & 0.04 & 24\% & 35\% \\
      $1.0<z<1.2$ & 0.05 & 38\% & 10-15\%\\
      $1.2<z<1.4$ & 0.1 &  79\% & 4\% \\
      $1.4<z<2$ & None & 100\% & None \\
      \hline
\end{tabular}    \vspace{2mm}
\end{center}
 
 \begin{center}
    \begin{tabular}{|l|c|c|c|}
     \multicolumn{4}{c}{\textbf{DES $grizY$ + VHS $JHK_s$ photometry}} \\ 
    \hline
  Redshift Range & Optimum Clipping Threshold & Fraction of galaxies used & Improvement in $\delta P/P$ \\ 
      \hline
      $0<z<1$ & 0.03 & 8\% & 30\%\\
      $1<z<1.3$ & 0.05 & 37\% & 10-15\% \\
      $1.3<z<2$ & 0.1 & 80\% & 15-20\% \\
      \hline
 \end{tabular}    \vspace{2mm}
  \end{center}
\caption{Summary of the optimum threshold error to be applied in different redshift ranges in order to minimise the fractional error on the power spectrum. The top table is for DES $grizY$ photometry and the bottom table for DES+VHS $JHK_s$ photometry.               \label{tab:deltaP}} 
\end{table*}

From these results we can clearly see that there exists a trade-off between the shot-noise contribution to the error on the power spectrum and the contribution from cosmic variance. At high values of the threshold error, most of the galaxies in the sample are used for analysis and shot-noise is not a problem. However, the scatter on the photometric redshift is large leading to larger errors in the power spectrum measurement. At very low values of the threshold error, the photo-z scatter is reduced but there are too few galaxies in the sample and shot noise begins to dominate. There is an optimum value of the threshold error at which the fractional errors in the power spectrum are at a minimum. This value is different for different redshift ranges as well as for the two different catalogues. 

When we add the VHS NIR data to the DES optical catalogue, we can apply a smaller clipping threshold out to the same redshift range compared to the DES only case in order to minimise the error in the power spectrum. This means we remove more outliers from the DES+VHS catalogue and thereby reduce our photometric redshift errors without compromising on the precision with which we can do cosmology. Also, we can see that using the DES only catalogue, we are unable to clip in the highest redshift bin of $1.4<z<2$ as this increases the shot noise errors in our power spectrum. However, if we add the VHS NIR photometry, we can remove $\sim$20\% of our galaxies in this bin and produce a power spectrum that is 15-20\% more accurate than that obtained using all the galaxies.  

We can conclude that in the absence of large spectroscopic surveys like the proposed WFMOS survey \citep{Bassett:WFMOS}, photometric surveys could prove competitive in constraining dark energy through galaxy power spectrum measurements if the outliers were effectively removed.  

It is worth noting though that when applying this clipping procedure to a real survey, one would choose the optimal clipping threshold based on the training sets available and not from simulations. 

\subsection{Effect of Photometric Redshift Bias on Dark Energy Equation of State}

\label{sec:cosmology2}

In this section we look at the effect of the photometric redshift bias on the dark energy equation of state parameter, $w$. We have already seen in $\S$\ref{sec:bias} that systematic errors in the photometric redshift bias can arise when we use different numbers of galaxies in our training and testing sets and also when one or both of these samples is in some way incomplete. We can translate the errors in the bias given by the right hand panels of Figure \ref{fig:nv} and Figure \ref{fig:biases} into an error on the the value of $w$ calculated using baryon acoustic oscillations as a probe. The position of the BAO peaks can be used to find the angular diameter distance, $D_A$ which in turn tells us about the expansion history of the Universe and hence $w$. If there is a systematic uncertainty on the photometric redshift bias, $\Delta_{b}$, this can be related to the uncertainty in $w$ using the angular diameter distance, in the following way:

\begin{equation}
\Delta w=\frac{\partial D_A}{\partial z}\frac{\partial w}{\partial D_A}\Delta_{b}
\label{eq:dw}
\end{equation} 

\begin{figure}
\begin{center}
\includegraphics[width=8.5cm,height=7.5cm,angle=0]{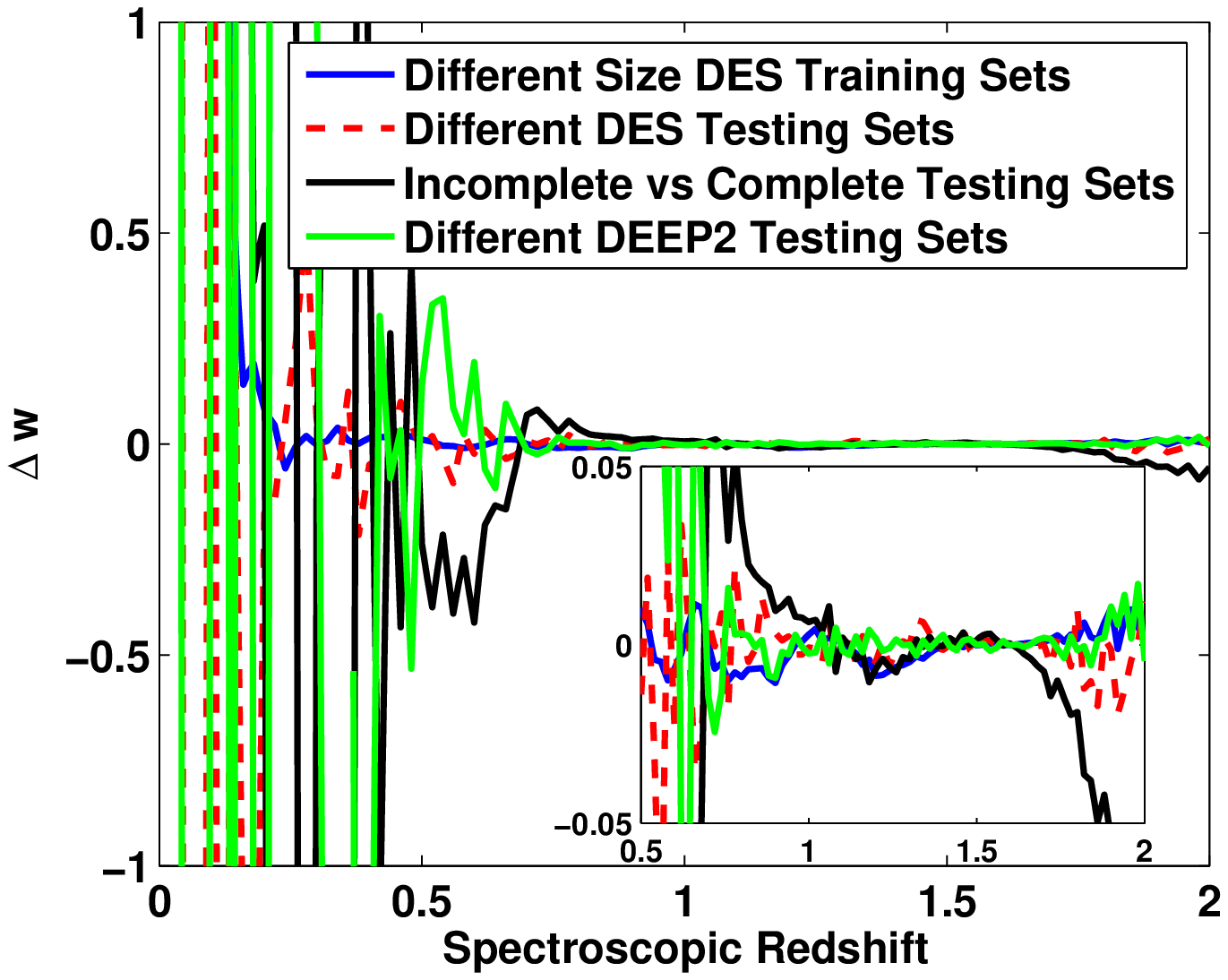}
\caption{The error in $w$ as a function of redshift due to the photometric redshift bias.}
\label{fig:dw}
\end{center}
\end{figure}

By assuming that $\Delta_{b}$ is given by the difference curves plotted in the right hand panels of  Figure \ref{fig:nv} and Figure \ref{fig:biases}, we can find $\Delta w$ for each of the cases investigated in $\S$\ref{sec:bias}. This is shown in Figure \ref{fig:dw}. Note that throughout this calculation we keep all other cosmological parameters constant and use a standard cosmology with $\Omega_m$=0.3, $\Omega_{\Lambda}$=0.7 and $h$=0.7.  

\begin{figure}
\begin{center}
\includegraphics[width=8.5cm,height=7.5cm,angle=0]{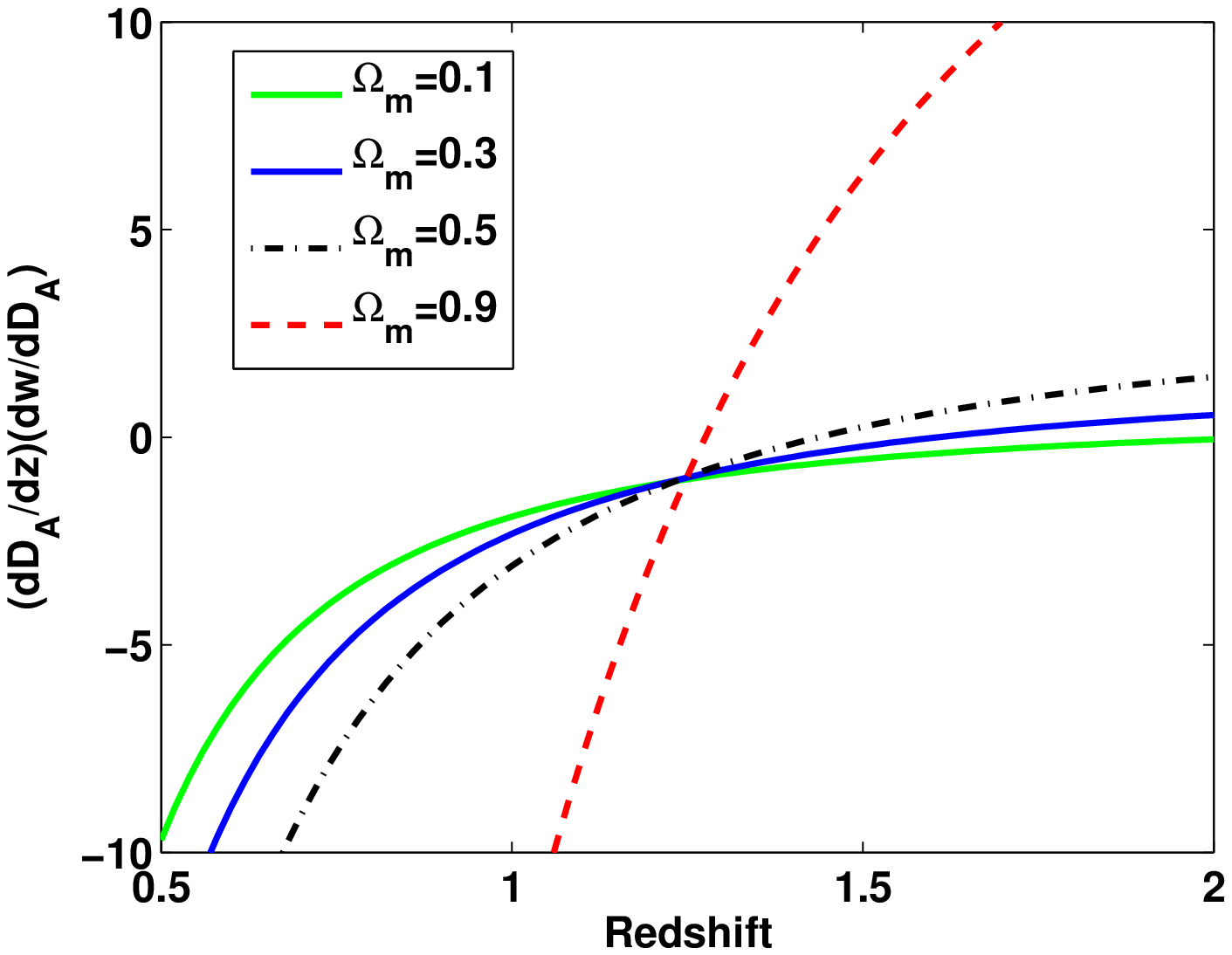}
\caption{$\frac{\partial D_A}{\partial z}\frac{\partial w}{\partial D_A}$ for four different cosmologies with $\Omega_m=0.1$, $\Omega_m=0.3$, $\Omega_m=0.5$ and $\Omega_m=0.9$. All models assume a flat universe.}
\label{fig:deriv}
\end{center}
\end{figure}

For interest, we also plot the product of the two derivatives in Eq. \ref{eq:dw} for four different cosmologies in Figure \ref{fig:deriv} as this is what links the error in the photometric redshift to the error on $w$. We assume a flat universe and change $\Omega_m$ in order to get four different cosmologies with different amounts of matter and dark energy. 

We can see that using different size training sets or different testing sets leads to an error in $w$ at a given redshift \footnote{This does not correspond to the error one would get from fitting a constant $w$ but rather the error of measuring $w$ at that given redshift.} that is of the order of $0.01$ for $z>0.5$. The error in $w$ is of the order of $0.08$ for $0.5<z<1$ and $z>1.5$ when we use an incomplete testing set with an incomplete training set as opposed to a complete testing set with an incomplete training set. The error is smaller - around $0.01$ in the redshift range $1<z<1.5$. The error in all cases is very large at $z<0.5$. As using different training and testing sets is equivalent to having a cosmic variance limited sample, we conclude that systematics on the photo-z bias due to cosmic variance do not hinder the calibration of $w$ to a percent level. However, the black solid line of Figure \ref{fig:dw} shows us that using an incomplete training set to calibrate photometric redshifts for DES would lead to large uncertainties in $w$ unless the sample being tested was cut to match the training set. These results are relatively insensitive to the clipping procedure introduced in $\S$ \ref{sec:clip} and therefore have been carried out in all cases for the umclipped catalogues of galaxies. 

Although this is a rough calculation, it gives us a feel for the uncertainties in calculating the dark energy equation of state that can arise due to the differences in the photometric redshift bias. In order to get the exact uncertainty on $w$ due to uncertainties in the photometric redshift, one would have to conduct a full Fisher matrix analysis applied to baryon acoustic oscillations such as that done by \citet{Ma:photoz_wl} for weak lensing. This involves translating the error on the bias and scatter in the photometric redshift to an error on $w$ by marginalising over all other cosmological parameters.

From this work, we can conclude that although we can use neural networks reasonably successfully to obtain photometric redshifts by extrapolating from an incomplete training set, if this is done with a survey such as DES it will create systematic errors on $w$ of the order of $\sim$10\%. The possibility of using a template-fitting method to calculate photometric redshifts in regions where the training set is incomplete should be further investigated.    

\section{Conclusions}

In this work we have shown the role of near infra-red photometry from the VISTA Hemisphere Survey in constraining photometric redshifts for the Dark Energy Survey. We have examined the effects of galaxy reddening and training sets on the photo-z estimate. We have also studied the biases in the photometric redshift estimate when using training and testing sets with different sizes and levels of completeness and quantified the error in the dark energy equation of state parameter, $w$ that arises from differences in these biases.  

A method of clipping our galaxy catalogues by removing outliers based on the ANNz error estimate on the photometric redshift has been presented. By applying a threshold error to our catalogues and rejecting all objects with errors bigger than this threshold error, we can effectively reduce the overall scatter on the photometric redshifts of our sample.  

Finally, we have conducted a full galaxy power spectrum analysis using our DES and DES+VHS catalogues and looked at how our clipping method can improve the uncertainties on our galaxy power spectrum measurements. We find that there is an optimum threshold error at which we should clip our catalogues and this error depends on the catalogue being used and the redshift range in which we are evaluating the power spectrum. If we use a high value for the threshold error, the scatter on our photometric redshift estimate is high leading to large positional uncertainties and therefore large errors in the power spectrum due to cosmic variance. However, if we adopt a very low value for our threshold error, we remove most of the galaxies from our sample before calculating the power spectrum and the resulting uncertainties in the power spectrum are dominated by shot noise. We find that the optimum threshold error is smaller for the DES+VHS catalogues compared to the DES only catalogues in the same redshift range and hence more outliers are removed from this sample before analysis. Adding the VHS NIR data thus helps us to compute the galaxy power spectrum more accurately out to higher redshifts than for the DES only case.

In summary, our main conclusions are:

\begin{itemize}

\item{NIR data from VISTA VHS helps to reduce the scatter on DES photometric redshifts by $\sim$30\% for $z>1$.}

\item{Reddening the galaxies can increase the photo-z scatter of DES by $\sim$ 30\% in some redshift ranges due to the degeneracy between redshift and reddening that exists in these redshift ranges. However, this is unlikely to be a major issue as most of our mock DES galaxies do not suffer from heavy extinction.}

\item{ANNz can be used to predict the extinction, $A_v$ of DES galaxies to an accuracy of 0.27 and to classify them into six spectral types - E, Sbc, Scd, Im, SB2 and SB3.}

\item{The VVDS-Deep and DEEP2 spectroscopic surveys, when finished will provide a very complete training set for DES out to a redshift of 2.}

\item{Using different numbers of training set galaxies can lead to a difference in the photometric redshift bias of the order of $10^{-3}$.} 

\item{If we have an incomplete training set, we can improve the photometric redshift estimate by imposing the same colour cuts on the testing set as are applied to the training set. When this is done, the improvement in the photometric redshift bias is of the order of 10\% compared to if we used a complete testing set with no imposed colour cuts.}

\item{The clipping method introduced by \citet{Abdalla:DUNEphotoz} can be effectively applied to the DES+VHS sample and applying a threshold error of 0.1 at which to cut our sample, reduces the scatter on the photometric redshift by $\sim$50\% by removing $\sim20$\% of the galaxies.}

\item{A clipping threshold of 0.1 is optimal for calculating the DES power spectrum out to a redshift of 1.4. Applying this clipping threshold reduces $\delta P/P$ by $\sim$20\%. When the VHS NIR data is added to the DES sample, the optimal clipping threshold in the same redshift range is 0.05 and this reduces the fractional error in the power spectrum by $\sim$20-30\%. In order to calculate the power spectrum out to a redshift of 2, the addition of VHS NIR data is crucial. In this redshift range, applying the optimal clipping threshold of 0.1 results in an improvement in $\delta P/P$ by $\sim$15-20\%.}

\item{Systematic errors on the photometric redshift bias arising from cosmic variance lead to uncertainties in the dark energy equation of state parameter, $w$, of about a percent.}

\item{However if we use an incomplete training set to determine photometric redshifts on a testing set that hasn't been cut to match the training set, the resulting uncertainties in the photometric redshift bias can lead to errors in $w$ of the order of $\sim$10\% if we keep all other cosmological parameters as constant. Note however that this result is model specific and depends to a certain extent on our choice of mock catalogues and the algorithm used to calculate photometric redshifts, in this case, the neural networks.}

\end{itemize}

In the absence of large spectroscopic surveys, the DES and VHS datasets, when combined, will prove extremely effective in constraining dark energy through large scale structure signals like baryon acoustic oscillations. By clipping photometric redshift catalogues and carefully removing a suitable number of outliers, one can achieve reasonably precise measurements of the galaxy power spectrum out to a redshift of 2.  

\section*{Acknowledgements}

We are very grateful to Peter Capak for providing the JPLCAT simulations. We thank members of the DES photometric redshift and large scale structure working groups for useful discussions, in particular Josh Friemann, Enrique Gaztanaga and Will Percival. We also thank Richard McMahon and Will Sutherland for information regarding the VISTA public surveys. MB is supported by an STFC studentship. FBA acknowledges support from the Leverhulme Foundation through an Early Careers Fellowship.  


\bibliography{}

\end{document}